\documentclass{jfm}

\usepackage{graphicx}
\usepackage{natbib}
\usepackage{color}
\usepackage{amsfonts,amssymb,amsmath}
\usepackage{bm}

\graphicspath{{./}}

\ifCUPmtlplainloaded \else
  \checkfont{eurm10}
  \iffontfound
    \IfFileExists{upmath.sty}
      {\typeout{^^JFound AMS Euler Roman fonts on the system,
                   using the 'upmath' package.^^J}%
       \usepackage{upmath}}
      {\typeout{^^JFound AMS Euler Roman fonts on the system, but you
                   dont seem to have the}%
       \typeout{'upmath' package installed. JFM.cls can take advantage
                 of these fonts,^^Jif you use 'upmath' package.^^J}%
      }
  \else
  \fi
\fi

\ifCUPmtlplainloaded \else
  \checkfont{msam10}
  \iffontfound
    \IfFileExists{amssymb.sty}
      {\typeout{^^JFound AMS Symbol fonts on the system, using the
                'amssymb' package.^^J}%
       \usepackage{amssymb}%

      }{}
  \fi
\fi

\ifCUPmtlplainloaded \else
  \IfFileExists{amsbsy.sty}
    {\typeout{^^JFound the 'amsbsy' package on the system, using it.^^J}%
     \usepackage{amsbsy}}
    {\providecommand\boldsymbol[1]{\mbox{\boldmath $##1$}}}
\fi

\newsavebox{\astrutbox}
\sbox{\astrutbox}{\rule[-5pt]{0pt}{20pt}}

\newcommand{\BS}[1]{{\color{black}#1}}
\newcommand{\SB}[1]{{\color{black}#1}}

\title[trigger turbulent bands in channel flow]{
Trigger turbulent bands directly at low Reynolds numbers in channel flow using a moving-force technique
}

\author[Baofang Song, Xiangkai Xiao]
{Baofang Song 
\thanks{Email address for correspondence: baofang\_song@tju.edu.cn}
, Xiangkai Xiao}
\affiliation{Center for Applied Mathematics, Tianjin University, Tianjin 300072, China 
}

\date{?; revised ?; accepted ?. - To be entered by editorial office}
\begin{document}

\maketitle

\begin{abstract}
We show a method, for direct numerical simulations, to trigger and maintain turbulent bands directly at low Reynolds numbers in channel flow. The key is to impose a moving localised force which induces a local flow with sufficiently strong 
inflectional instability. With the method, we can trigger and maintain turbulent bands at Reynolds numbers down to $Re\simeq 500$. More importantly, we can generate any band patterns with desired relative position and orientation. The usual perturbation approach resorts to turbulent fields simulated at higher Reynolds numbers, random noise, or localised vortical perturbation, which neither assures a successful generation of bands at low Reynolds numbers nor offers a control on the orientation of the generated bands. \BS{A precise control on the position and orientation of turbulent bands is important for the investigation of all possible types of band interaction and for understanding the transition in channel flow at low Reynolds numbers.}

\end{abstract}

\begin{keywords}
\end{keywords}

\section{Introduction}
Since the work of \citet{Tsukahara2005}, it has been known that turbulence in channel flow appears in form of turbulent bands at low Reynolds numbers, which are tilted with respect to the streamwise direction \citep{TsukaharaKawaguchi2014,TsukaharaKawamura2014,Tuckerman2014,Xiong2015,Tao2018,Kanazawa2018,Shimizu2019,Xiao2020,Paranjape2020}. Similar band patterns were also observed in other types of shear flows at transitional Reynolds numbers \citep{Coles1965,Prigent2002,Barkley2005,Duguet2010,Tuckerman2011,Duguet2013,Rolland2015,Rolland2016,Chantry2017,Reetz2019,Tuckerman2020}. 
\SB{The obliqueness of turbulent bands and the sustaining mechanism of bands in tilted periodic domains have been explained from the point of view of the large scale flow and dynamical system approach \citep{Duguet2013,Reetz2019,Paranjape2020}.  However, the mechanism underlying the specific tilt angles and the dynamcics of fully localised turbulent bands remain poorly understood.} Latest numerical studies showed that individual turbulent bands in large channels can be sustained at $Re\gtrsim 660$, and once triggered, a turbulent band can grow transversely or split when the band length is sufficiently large \citep{Xiong2015,Tao2018,Shimizu2019,Kanazawa2018}. Throughout this paper, \BS{constant-flux driven flow is discussed about and} the Reynolds number is defined as $Re=U_ch/\nu$, where $U_c$ is the centerline velocity of the unperturbed parabolic flow, $h$ the half gap-width and $\nu$ the kinematic viscosity of the fluid. 

Regarding the sustaining mechanism of fully localised turbulent bands at low Reynolds numbers, \citet{Shimizu2019,Kanazawa2018} observed that turbulent bands are sustained by an active streak-generating downstream end (hereafter referred to as head). \citet{Kanazawa2018} considered a damped Navier-Stokes system and proposed that a relative periodic orbit seems to exist at the head and is responsible for the self-sustainment of the band. \citet{Xiao2020} lately investigated the local mean flow at the head of turbulent bands at $Re=750$ and showed that the spanwise velocity profile manifests a strong inflection. They performed linear stability analysis of the inflectional profile and found a linear instability. The most unstable eigenmode as well as the nonlinear development of perturbations on top of the inflectional velocity profile showed remarkable similarities with the wave-like streaky structure observed at the head of turbulent bands. They also showed that streaks decay continually at the upstream end of turbulent bands (hereafter referred to as tail). Therefore, they proposed that the growth of fully localised turbulent bands are driven by the instability associated with the inflectional local mean flow at the head. Similar mechanisms were proposed for the growth mechanism of the wing tips of turbulent spots at higher Reynolds numbers in channel flow \citep{Henningson1987,Henningson1989}, in plane Couette flow \citep{Dauchot1995} and at the laminar-turbulent interface of turbulent puffs in pipe flow \citep{Hof2010}. \SB{However, how this inflectional flow is formed and sustained and the possible connection to the periodic orbit solution of \citet{Kanazawa2018} remain to be investigated.} 
 
In the presence of multiple turbulent bands, interactions between bands may occur and result in complex spatio-temporal intermittency \citep{Duguet2010,Shimizu2019}. Interactions between turbulent bands in channel flow at low Reynolds numbers have not been \BS{sufficiently investigated}. It was shown that, in small tilted domains, turbulence forms parallel band patterns \citep{Tuckerman2014}; however, the choice of the size and tilt direction of the domain restrict the turbulent bands from taking different orientations. In non-tilted domains, \citet{Tao2018} reported that, if the domain is not sufficiently large, a band may decay due to the self-interaction because of the periodic boundary conditions. They argued that this is because the large scale flow around a band is necessary for its self-sustainment, and the large scale flow may be affected by itself in a periodic channel or potentially by a close neighbouring band. Yet they did not explicitly study interactions between turbulent bands, especially bands with different orientations. To our knowledge, in numerical simulations, so far only \citet{Shimizu2019} investigated interactions between multiple bands in a large computational domain ($500h\times 2h\times 250h$ in the streamwise, wall normal and spanwise directions, respectively). The large domain with periodic boundary conditions allowed them to observe the development of and the interactions between multiple bands for very long times, which were $\mathcal{O}(10^5)$ convective time units and very much out of reach of current laboratory experiments. They concluded that at $Re\gtrsim 976$ (higher than 830 as proposed by \citet{Sano2016}), directed percolation (DP) model can be used to model the interactions between bands and the resulting flow pattern. Before the onset of DP, turbulent bands can still be sustained. \BS{They observed longitudinal interaction between parallel bands and collision between bands with opposite orientations, and that turbulent bands can only form parallel pattern instead of two-sided pattern as they do at higher Reynolds numbers. However, more quantitative studies on the interactions between turbulent bands in this flow regime are needed to elucidate the transition scenario at low Reynolds numbers.} 

Considering the fact that turbulent bands can take opposite orientations (angles about the streamwise direction), it is necessary to generate bands with desired positions and orientations in order to study all possible types of interactions between bands. In numerical simulations, to our knowledge, usually two methods have been used to generate bands at low Reynolds numbers ($Re\lesssim 1000$). The one is to start from a fully turbulent flow field at a higher Reynolds number (above $Re\simeq 1500$) or random noise, and wait for the flow to form discrete bands \citep{Tsukahara2005,TsukaharaTakeda2014,Tuckerman2014}. The other is to start from a localised vortical perturbation such as that proposed by \citet{Henningson1991}. As \citet{Tao2018} pointed out that random noise cannot trigger turbulent bands at $Re<800$, no matter being localised or not, and that only perturbations with effective structures can work. In our simulations, we also found that both methods cannot assure the generation of bands at $Re\lesssim 750$, regardless of the amplitude of the initial perturbations. At higher Reynolds numbers, even if bands can be triggered by noisy fields or localised vortical perturbations, the tilt direction of the bands are not known a priori using both methods and the first method even cannot predict the number and positions of the generated bands.

In this work, we propose a method that can effectively trigger turbulent bands directly at low Reynolds numbers and more importantly, allows a precise control on the position and orientation of the generated bands. 

\section{Methods}
The non-dimensional incompressible Navier-Stokes equations
\begin{equation}\label{equ:NS}
 \frac{\partial \bm u}{\partial t}+{\bm u}\cdot\bm{\nabla}
{\bm u}=-{\bm{\nabla}p}+\frac{1}{Re}{\bm\nabla^2}{\bm u} + {\bm F}, \;
\bm{\nabla}\cdot{\bm u}=0
\end{equation}
with a constant volume flux for the channel geometry are solved in Cartesian coordinates $(x, y, z)$, where $\bm u$ denotes velocity, $p$ denotes pressure, $\bm F$ denotes the external force and $x$, $y$ and $z$ represent the streamwise, wall-normal and spanwise coordinates, respectively. Velocities are normalized by $U_c$, length by $h$ and time by $h/U_c$. \BS{For all simulations in this paper, the volume flux associated with the unperturbed parabolic flow is imposed.} No-slip boundary conditions for velocities are imposed at channel walls (i.e. at $y=\pm 1$).  Periodic boundary conditions are imposed in the streamwise and spanwise directions. A hybrid Fourier spectral-finite difference method is used to solve Eqs. (\ref{equ:NS}), with a finite-difference method with a 9-point stencil employed for the discretisation in the wall normal direction. Therefore, the velocity and pressure fields can be expressed as
\begin{equation}
{A}(x,y,z,t)=\sum_{k=-K}^{K}\sum_{m=-M}^{M}\hat{A}_{k,m}(y,t)e^{i(\alpha kx+\beta mz)},
\end{equation}
where $k$ and $m$ are the indices of the streamwise and spanwise Fourier modes, respectively, $\hat A_{k,m}$ is the Fourier coefficient of the mode $(k,m)$ and $\alpha$ and $\beta$ are the fundamental wave numbers in the streamwise and spanwise directions, respectively. The size of the computational domain is $L_x=2\pi/\alpha$ and $L_z=2\pi/\beta$. The finite difference scheme and the parallelisation strategy of {\sc Openpipeflow} \citep{Willis2017}, and the time-stepping and projection scheme of \citet{Hugues1998} are employed to integrate the incompressible system.

We performed direct numerical simulations (DNS) in the regime of $Re\leqslant 750$. We considered two computational boxes with $L_x=L_z=120h$ and $L_x=L_z=320h$, respectively. 
For $Re=750$, we used 768 Fourier modes ($K=M=384$) in both streamwise and spanwise directions for the small box and 2048 Fourier modes ($K=M=1024$) for the large box. For $Re=600$ and 500, the number of streamwise Fourier modes is reduced to 576 for the small box and 1728 for the large box. We used 64 Chebyshev grid points for the finite difference discretisation in the wall normal direction. These resolutions were shown to be sufficient for the flow in this Reynolds number regime \citep{Tao2018}. A time-step size of $\Delta t=0.01$ is used for the time integration, which is sufficiently small for this Reynolds number regime.

\subsection{The forcing}\label{sec:forcing}

\begin{figure}
\centering
\includegraphics[width=0.8\textwidth]{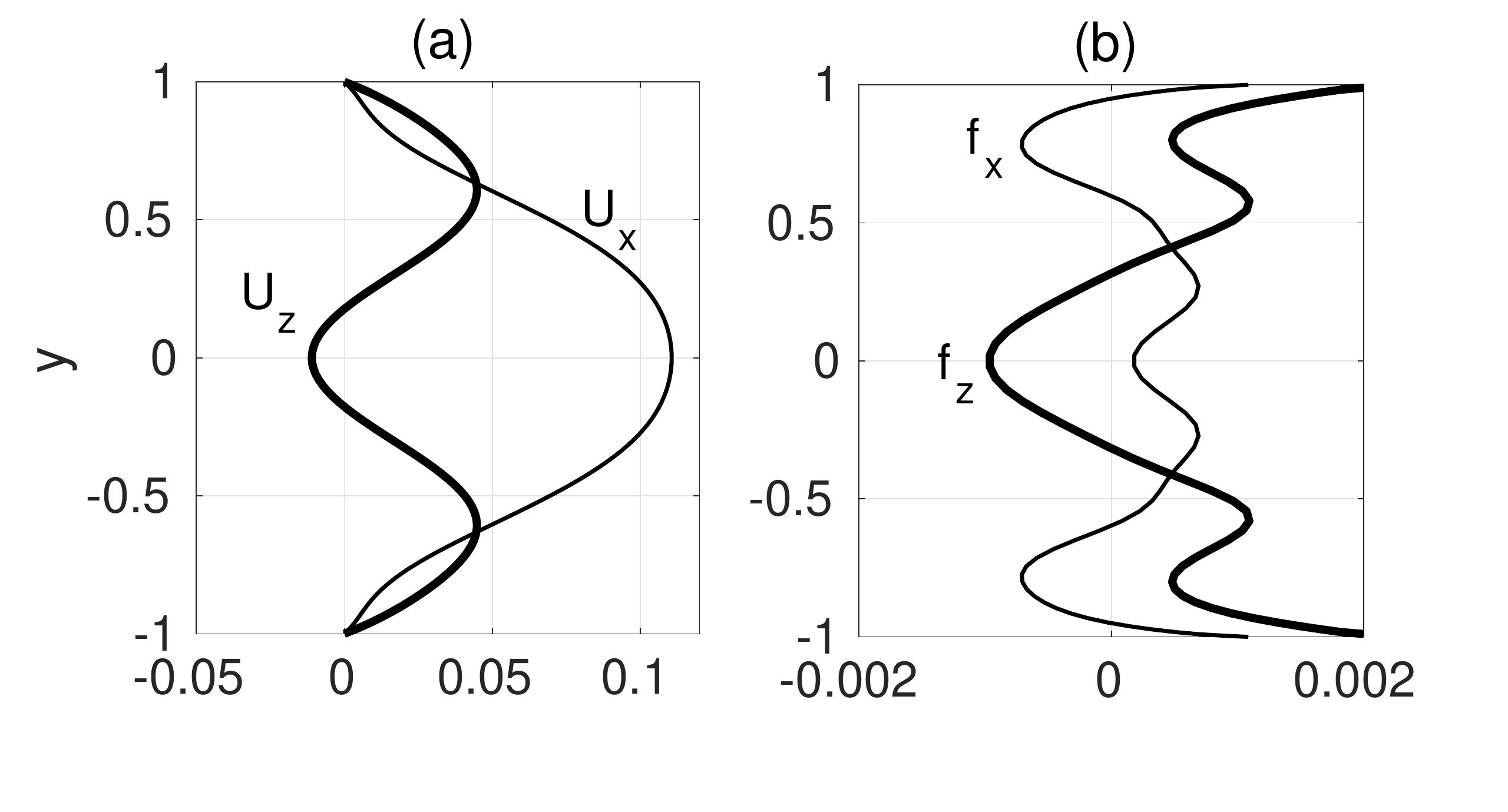}
\caption{\label{fig:target_profiles} (a) The velocity profiles (\ref{equ:target_profile}) (\ref{equ:target_profile_2}). (b) The forces that are determined by (\ref{equ:forcing}) using the velocity profiles in (a).}
\end{figure}

\citet{Xiao2020} showed that turbulent bands at low Reynolds numbers grow via a streak generation mechanism at the head of the bands, i.e. an inflectional instability associated with the local mean flow. Inspired by their study, here we aim to generate banded streaks, as those observed in turbulent bands, via such an instability mechanism. We propose to impose a body force that locally induces a spanwise inflectional velocity profile. We first design a target streamwise and spanwise velocity profile that is inflectionally unstable, and subsequently derive the force that can generate the profile. \BS{We use polynomial fits of the velocity profiles that \citet{Xiao2020} measured at the head of a turbulent band at $Re=750$, with the parabola $1-y^2$ subtracted from the streamwise component. The profiles read} 
\begin{eqnarray}
\label{equ:target_profile}
U_x&=&-0.2478y^8 + 0.5390y^6 -0.2768y^4-0.1250y^2+0.1106, \\
\label{equ:target_profile_2}
U_z&=&-0.2469y^8 + 0.7262y^6 -0.8448y^4+0.3765y^2-0.0110, 
\end{eqnarray}
and Figure \ref{fig:target_profiles}(a) shows the $y$-dependence of the profiles.
We derive the force $\boldsymbol f$ \BS{that is needed to maintain the profile} as a function of $y$ as
\begin{equation}\label{equ:forcing}
{\boldsymbol f}+\frac{1}{Re}\nabla^2 \boldsymbol U=0,
\end{equation}
\BS{where $\boldsymbol U=(U_x, 0, U_z)$. It should be noted that the force given by (\ref{equ:forcing}) is meant to generate velocity deviations $(U_x, 0, U_z)$ with respect to the parabola $1-y^2$. Figure \ref{fig:target_profiles}(b) shows the force determined by the target profiles.}

\begin{figure}
\centering
\includegraphics[width=0.7\textwidth]{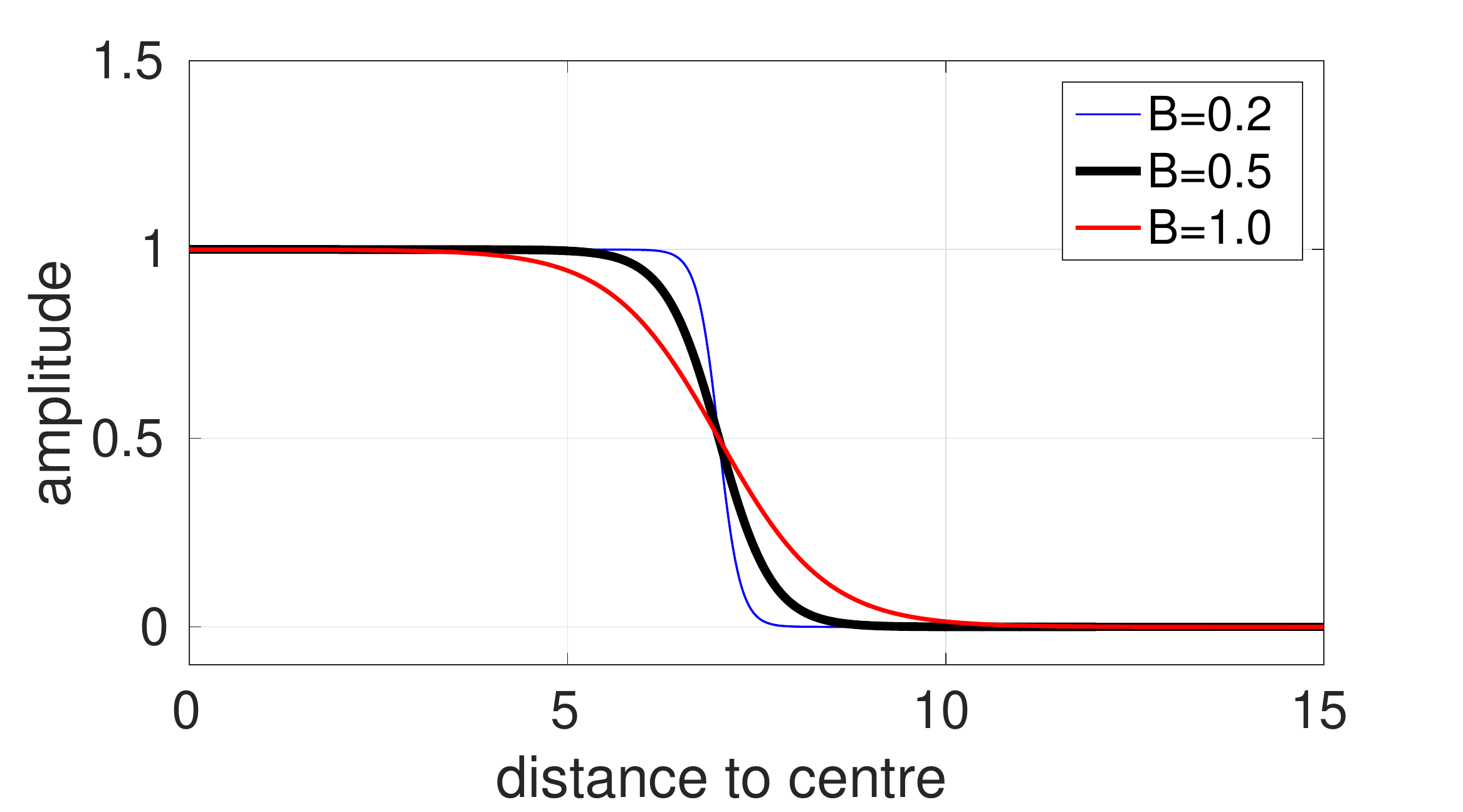}
\caption{\label{fig:localisation_factor} The localisation factor, i.e., the part in the parentheses in Eqs. \ref{equ:localised_forcing}, as a function of the distance to the centre of the forcing area, $\sqrt{(x-x_c)^2+(z-z_c)^2}$, given $R=7$ and $B=0.2$, 0.5 and 1.0.}
\end{figure}

Besides, to localise the force, we multiply the force by a tanh-function as the following
\begin{equation}\label{equ:localised_forcing}
\boldsymbol F=(F_x, 0, F_z)=\left(0.5-0.5{\text {tanh}} \frac{\sqrt{(x-x_c)^2+(z-z_c)^2}-R}{\sqrt{2} B}\right)(A_xf_x, 0, A_zf_z),
\end{equation}
where $B$ determines the sharpness of the localisation, $A_x$ and $A_z$ are the nominal amplitude of the force component $f_x$ and $f_z$, respectively, and are adjustable independently. $R$ is the nominal radius and $x_c$ and $z_c$ are the streamwise and spanwise coordinates of the center of the localised forcing area. As an example, figure \ref{fig:localisation_factor} shows the localisation factor as a function of the distance to the center of the forcing area in (\ref{equ:localised_forcing}) (the part in the parentheses) for $R=7$ and $B=0.2$, 0.5 and 1.0. The smaller $B$, the steeper the force curve at the boundary. The localisation factor falls nearly to zero at distances considerably larger than $R$ (e.g. larger than approximately 8 for $B=0.5$).

It should be noted that, in the proximity of the border of the forcing area, this localised force does not respect the incompressibility constraint, which is however fulfilled by the projection method when the Navier-Stokes equations are solved.
\BS{We want to point out that other velocity profiles with similar inflectional properties may be used to design the force.} 

\subsection{Stability property of the modified base flow}

\begin{figure}
\centering
\includegraphics[width=0.9\textwidth]{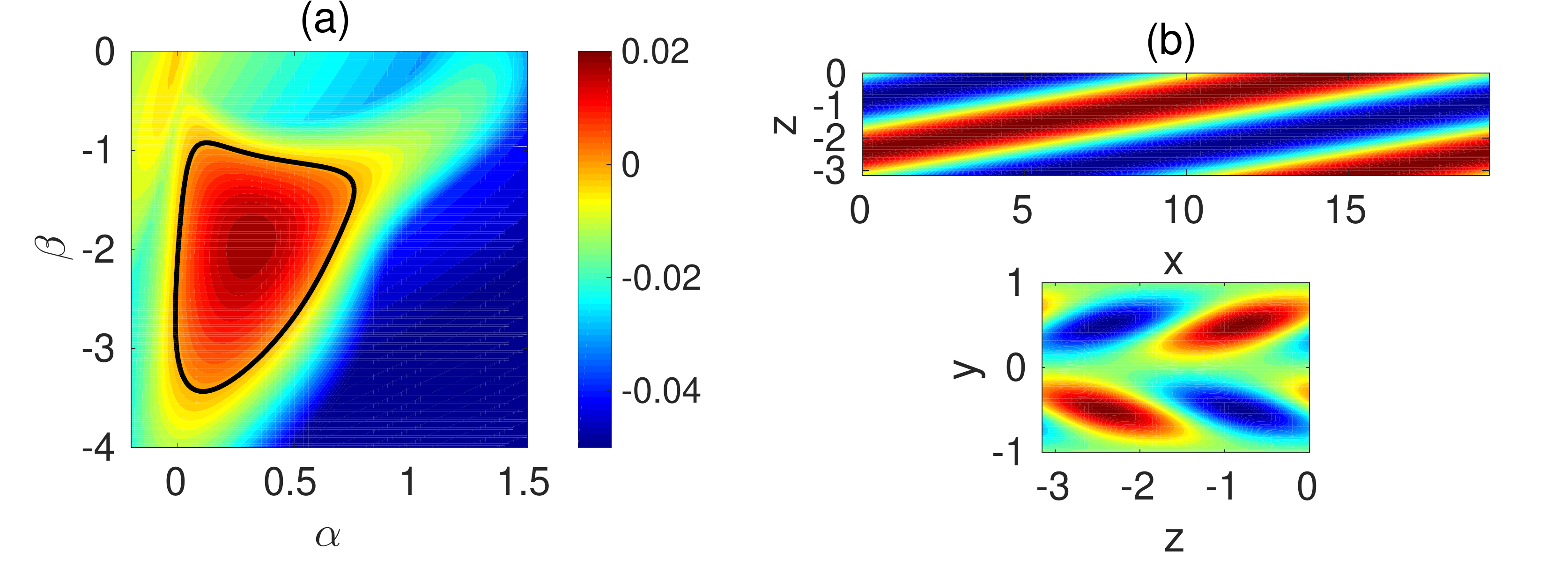}
\includegraphics[width=0.9\textwidth]{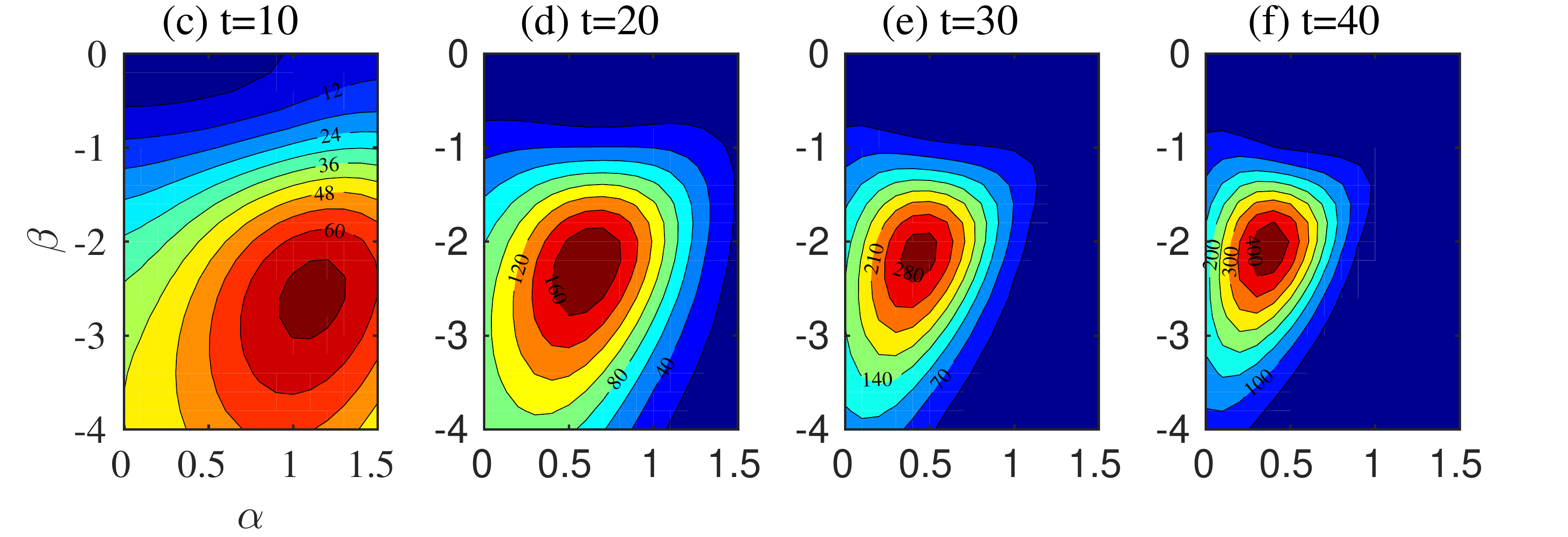}
\caption{\label{fig:unstable_region_and_nonmodal_growth} (a) Contours of the most unstable/least stable eigenvalue in the $\alpha$-$\beta$ wavenumber plane for the base flow $(U_x+1-y^2,0,U_z)$ at $Re=750$. The region circled by the bold black line marks the linearly unstable region. (b) The streamwise velocity of the most unstable mode $(\alpha,\beta)=(0.32,-1.96)$ is visualised in the $x$-$z$ cut plane at $y=-0.5$ (top) and in the $z$-$y$ cross-section at $x=0$ (bottom). The base flow is in the positive $x$ direction. Red color represents higher velocity and blue represents lower velocity compared to the base flow. (c-f) Contours of the non-modal energy growth in the $\alpha$-$\beta$ wavenumber plane at $t=10$, 20, 30 and 40.}
\end{figure}

Here, we briefly show the linear stability property of the modified base flow \BS{ $(U_x+1-y^2, 0, U_z)$}. \BS{The modal stability of this profile has been studied in \citet{Xiao2020}. For the ease of discussion, we reproduced the data in Figure \ref{fig:unstable_region_and_nonmodal_growth}(a, b).} The profile bears linear instabilities in a wide region in the wavenumber plane (the region enclosed by a bold line in panel (a)). The most unstable mode $(\alpha,\beta)=(0.32, -1.96)$ shows wave-like streaks tilted with respect to the streamwise direction, see the visualisation of the streamwise velocity in the $x$-$z$ cut plane at $y=-0.5$ (top) and in the $z$-$y$ cut plane (bottom) in panel (b). However, \citet{Xiao2020} mentioned that the modal growth rate of the most unstable mode seems smaller than the observed growth rate of the streaks at the head of turbulent bands. \BS{In this paper, we performed a non-modal analysis and will show that the unstable modes can actually achieve a significant and fast non-modal growth via the Orr-mechanism at the very early stage.} The adjoint-based method for non-modal energy growth described in \citet{Barkley2008} and \citet{Chai2019} is adopted for this analysis. 

The non-modal energy growth is shown as contours in the wavenumber plane at $t=10$, 20, 30 and 40; see Figure \ref{fig:unstable_region_and_nonmodal_growth}(c-f). It can be seen that at the very early stage, the most amplified mode is outside the linearly unstable region. Nevertheless, linearly unstable ones quickly become the most amplified modes after approximately $t=20$. Already at $t=30$, the linearly unstable modes are amplified by a few hundred times in energy.

\begin{figure}
\centering
\includegraphics[width=0.55\textwidth]{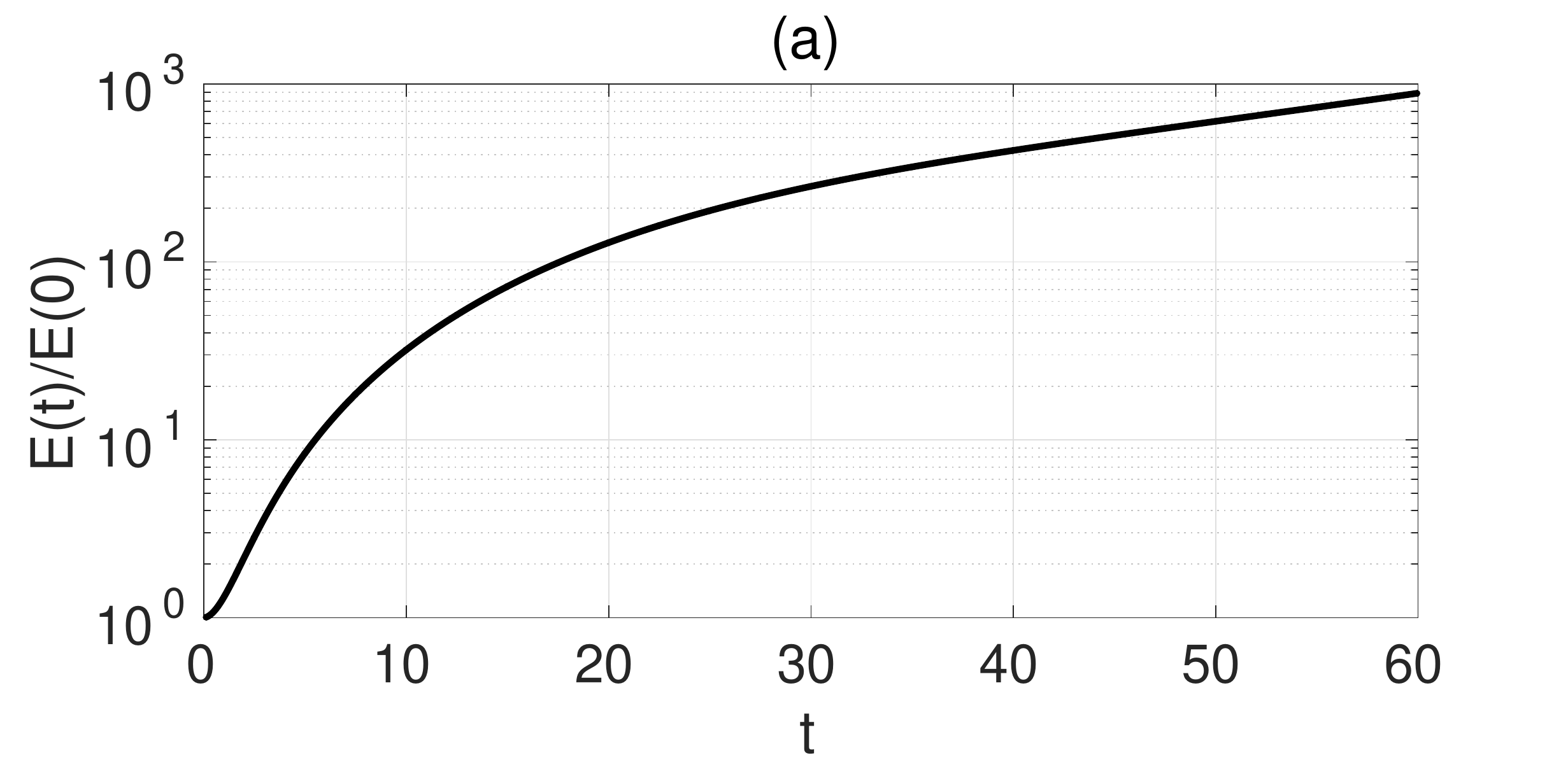}
\includegraphics[width=0.95\textwidth]{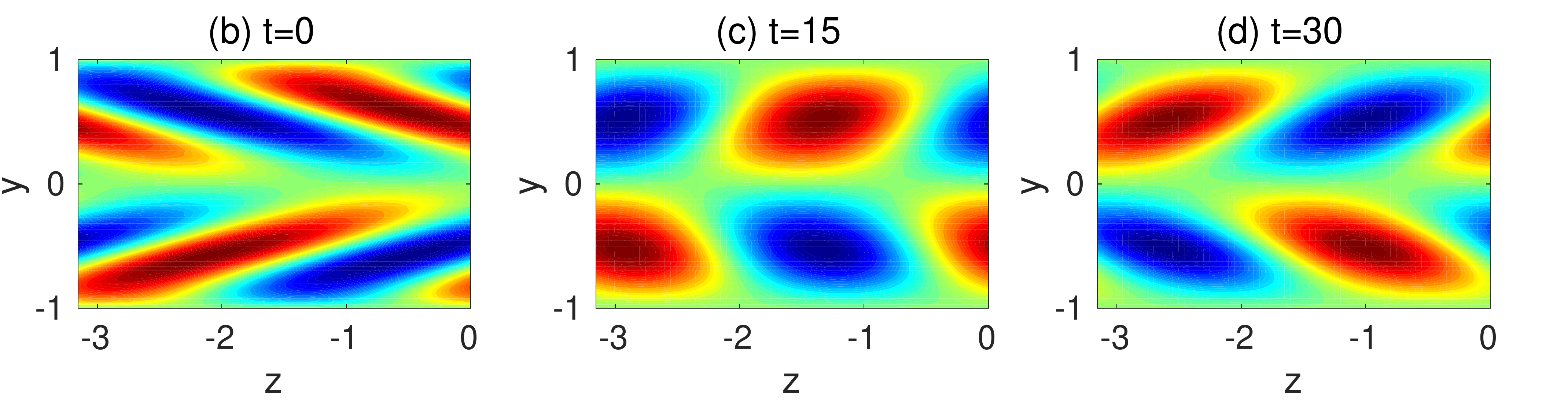}
\caption{\label{fig:non_modal_most_unstable} (a) The energy amplification $E(t)/E(0)$ of the optimal perturbation optimised at $t=30$ for the mode $(\alpha,\beta)=(0.32,-1.96)$, which is the most unstable mode of the modal instability analysis. $E(t)$ denotes the kinetic energy of the disturbances integrated in the whole domain at time $t$. (b-d) Change of the flow field of the flow described in panel (a). Contours of the streamwise velocity in the $z$-$y$ cut plane at $x=0$ is plotted at $t=0$, 15 and 30. Red color represents higher velocity and blue represents lower velocity compared to the base flow.}
\end{figure}
Figure \ref{fig:non_modal_most_unstable}(a) shows the energy amplification of the mode $(\alpha,\beta)$ = $(0.32,-1.96)$. The flow is initialised with the optimal perturbation that is optimised at $t=30$ in the non-modal analysis. The flow field is visualised in the $z$-$y$ cut plane at $t=0$, 15 and 30 in panel (b-d). As shown in panel (b), the initial flow field manifests flow structures that mainly tilt against the mean shear of the spanwise flow component (see Figure \ref{fig:target_profiles}(b)) in the region $-0.6\lesssim y\lesssim 0.6$. Quickly, only after about 15 time units, the tilt direction of the structures is inverted by the underlying mean shear, see panel (c), during which the perturbation energy is amplified by nearly 100 times, suggesting an Orr-mechanism. After $t\simeq 30$ the flow roughly takes the form of the most unstable mode that amplifies exponentially at later times (see panel (d) and panel (a)). The results indicate that small disturbances may achieve a large amplification of roughly 200 times within a short time via the Orr-mechanism and subsequently amplify exponentially due to the underlying modal instability. Jointly, these linear mechanisms can result in a fast and large energy growth of nearly 1000 times within about 60 time units (see Figure \ref{fig:non_modal_most_unstable}(a)). 

These properties of the spanwise inflectional velocity profile are very likely responsible for the fast streak generation at the head of turbulent bands at low Reynolds numbers. We use such velocity profiles to derive the force in order to induce local velocity profiles of similar stability properties and consequently trigger turbulent bands. 

\section{Generation of a single band}

\subsection{Sustained-band regime}
Here we show results at $Re=750$, which is in the regime of sustained turbulent band \citep{Tao2018,Kanazawa2018,Shimizu2019}. 
According to \citet{Xiao2020}, the head of a turbulent band at Re=750 travels at the speeds of $c_z=0.1$ (in absolute value) and $c_x=0.85$. Therefore, to mimic a moving head,  \SB{i.e. to maintain a moving (convective) local inflectional instability similar to that naturally sustained at the head of turbulent bands,} the forcing area is moved at the same speeds, i.e., $x_c=x_0 + c_x t$ and $z_c=z_0 + c_z t$, where $x_0$ and $z_0$ are the initial coordinates of the center of the forcing area. $R$ should be chosen such that the diameter of the forcing area is comparable with the streamwise width of turbulent bands at the head. The localisation parameter $B$ is chosen such that the forcing area is sufficiently localised and at the same time the force does not create a too steep speed variation at the boundary of the forcing area. In the results shown in this paper, we set $B=0.5$ (see Figure \ref{fig:localisation_factor}) unless explicitly stated. 

\begin{figure}
\centering
\includegraphics[width=0.9\textwidth]{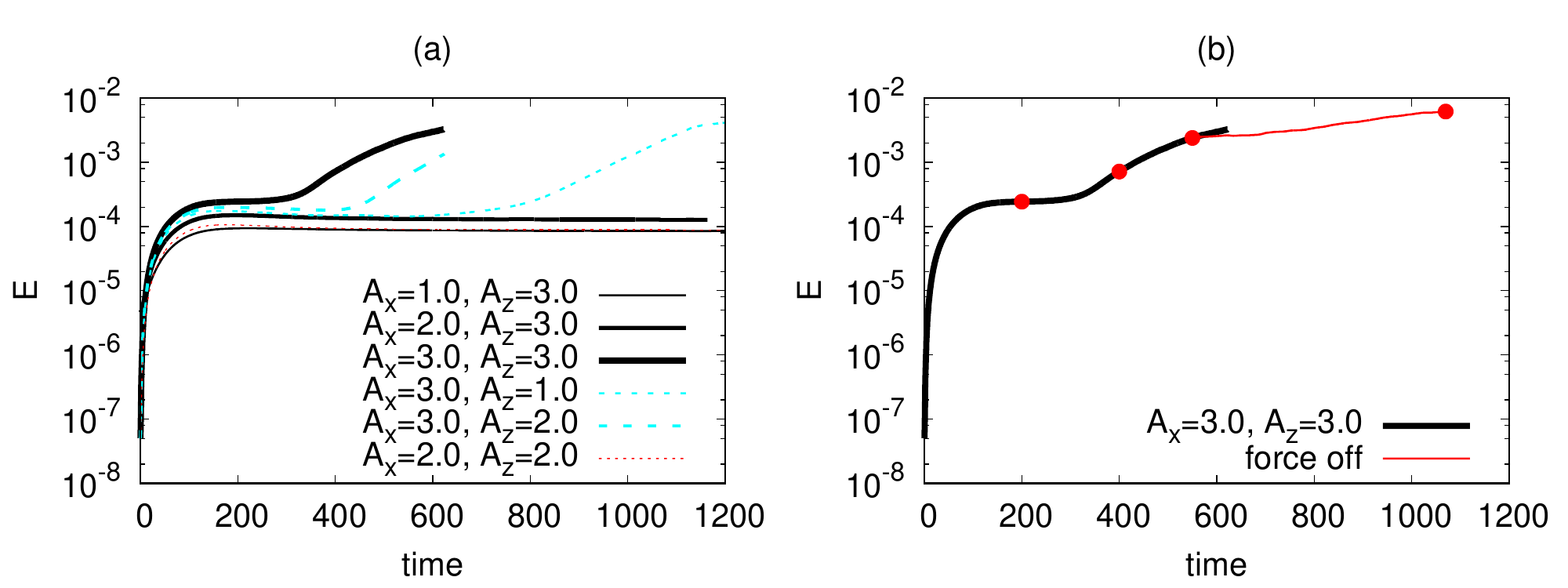}
\includegraphics[width=0.99\textwidth]{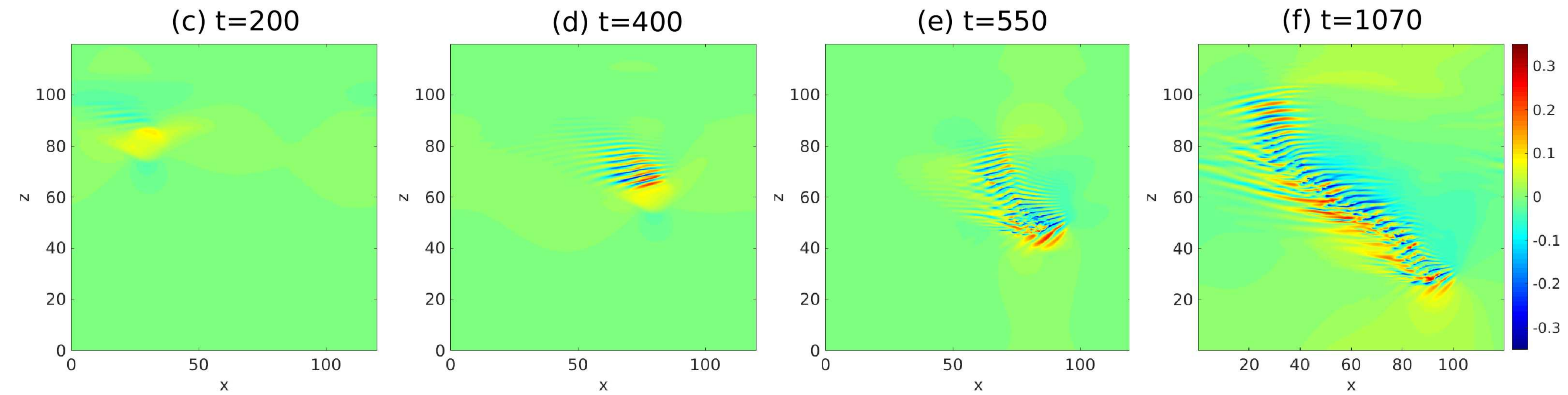}
\caption{\label{fig:forcing_amplitudes} (a) The effect of the forcing amplitude on the generation of bands at $Re=750$. The kinetic energy of the disturbances with respect to the parabola, integrated in the whole channel, is plotted. $A_x$ and $A_z$ denote the amplitude in the streamwise and spanwise directions, respectively. (b) The forces are switched off at $t=550$ and the following development of the flow is plotted as a thin red line. The case of $A_x=A_z=3.0$ is considered. (c-f) Contours of the streamwise velocity in the $x$-$z$ cut plane at $y=-0.5$. Time instants $t=200$, 400, 550 and 1070 for the $A_x=A_z=3.0$ case are shown. Flow is from left to right.}
\end{figure}

Firstly, we investigate the effects of the parameter $A$, i.e., the amplitude of the force. In order to mimic a noisy flow \BS{and to speed up the transition process}, the flow is initialised with a turbulent velocity field simulated at $Re=1500$, \BS{with the perturbations with respect to the parabola being} scaled by a factor of 0.001 (the resulting velocity perturbations are $\mathcal{O}(10^{-4})$. \BS{Note that a random noise applies equally}). The force is switched on at $t=0$. 
In the following tests, we set $R=7$ and set the amplitude of the force in the streamwise and spanwise directions, i.e. $A_x$ and $A_z$, independently. Figure \ref{fig:forcing_amplitudes}(a) shows the results. It can be seen that, for $A_x=3.0$, $A_z=1.0$ and 2.0 both fail to generate bands within 1200 time units, while $A_z=3.0$ succeeds (see the fast increase in the kinetic energy of disturbances after $t\simeq 300$). For $A_z=3.0$, $A_x=1.0$, 2.0 and 3.0 all successfully generate a turbulent band. The larger $A_x$ is, the earlier the turbulent band is generated. We also tested $A_x=A_z=2.0$ and found it fail to trigger a band. The results indicate that sufficiently strong forces are needed for the generation of turbulent bands and the stronger the force, the faster a band can be triggered, just as expected. The spanwise force, equivalently the spanwise velocity profile, seems to play a more important role than the streamwise one, which is consistent with our argument that the spanwise inflection plays a central role in the streak generation at the head. \SB{It should be noted that different velocity profiles (e.g. different $A_x$ and $A_z$ result in different velocity profiles) can be used to trigger instabilities and generate turbulent bands, as long as the profiles bear proper inflectional instability.}

For the case of $A_x=A_z=3.0$, we switched off the forcing at $t=550$ and the band generated by the forcing continues to grow, as indicated by the growing kinetic energy of the velocity field in Figure \ref{fig:forcing_amplitudes}(b). This indicates that the band generated by the forcing can be self-sustained even if the forcing is deactivated. Figure \ref{fig:forcing_amplitudes}(c-e) visualises the development of the flow under the forcing. We can see the locally deformed flow (see the yellow spot) at $t=200$ in panel (c). At $t=400$, tilted wavy structures (alternating high- and low-speed streaks) are generated by the instability of the forced local flow (see panel (d)) and the kinetic energy monotonically grows as more wavy streaks are generated. At $t=550$, a short band structure is generated with a streak-generating head at the downstream end (the right-bottom end) and a rather weak tail at the upstream end (the left-top end), which are the key characteristics of a turbulent band in channel flow at low Reynolds numbers \citep{Tao2018,Shimizu2019,Kanazawa2018}. After the force is deactivated, the band structure keeps growing and becomes self-sustained, indicated by the increasing kinetic energy of the flow. The flow at $t=1070$ is visualised in Figure \ref{fig:forcing_amplitudes}(f), which shows a well-developed turbulent band with a head, a tail and a wave-like elongated bulk.

\begin{figure}
\centering
\includegraphics[width=0.8\textwidth]{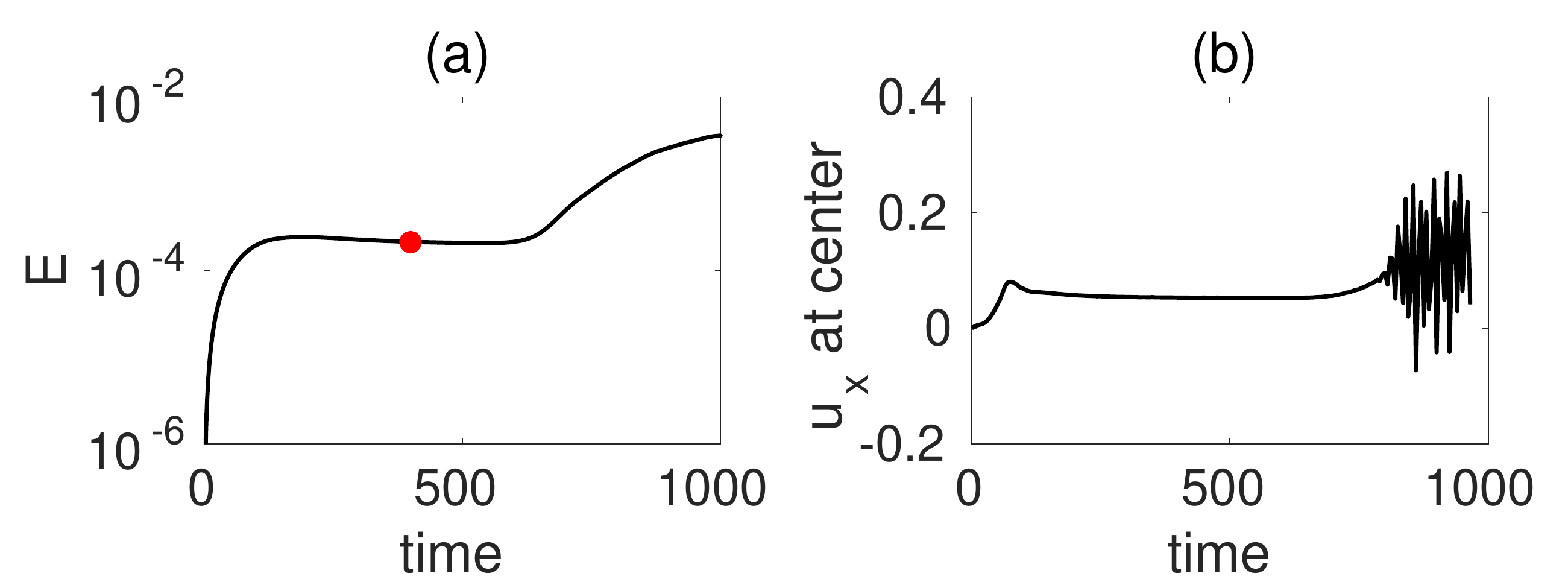}
\caption{\label{fig:KE_no_noise} \BS{(a) The kinetic energy of a forcing case without initial background noise. $Re=750$ and $A_x=A_z=3.0$. (b) The streamwise velocity $u_x$ (with the parabola excluded) at the center of the moving forcing area, monitored at the position of $y=-0.5$.}}
\end{figure}

\begin{figure}
\centering
\includegraphics[width=0.85\textwidth]{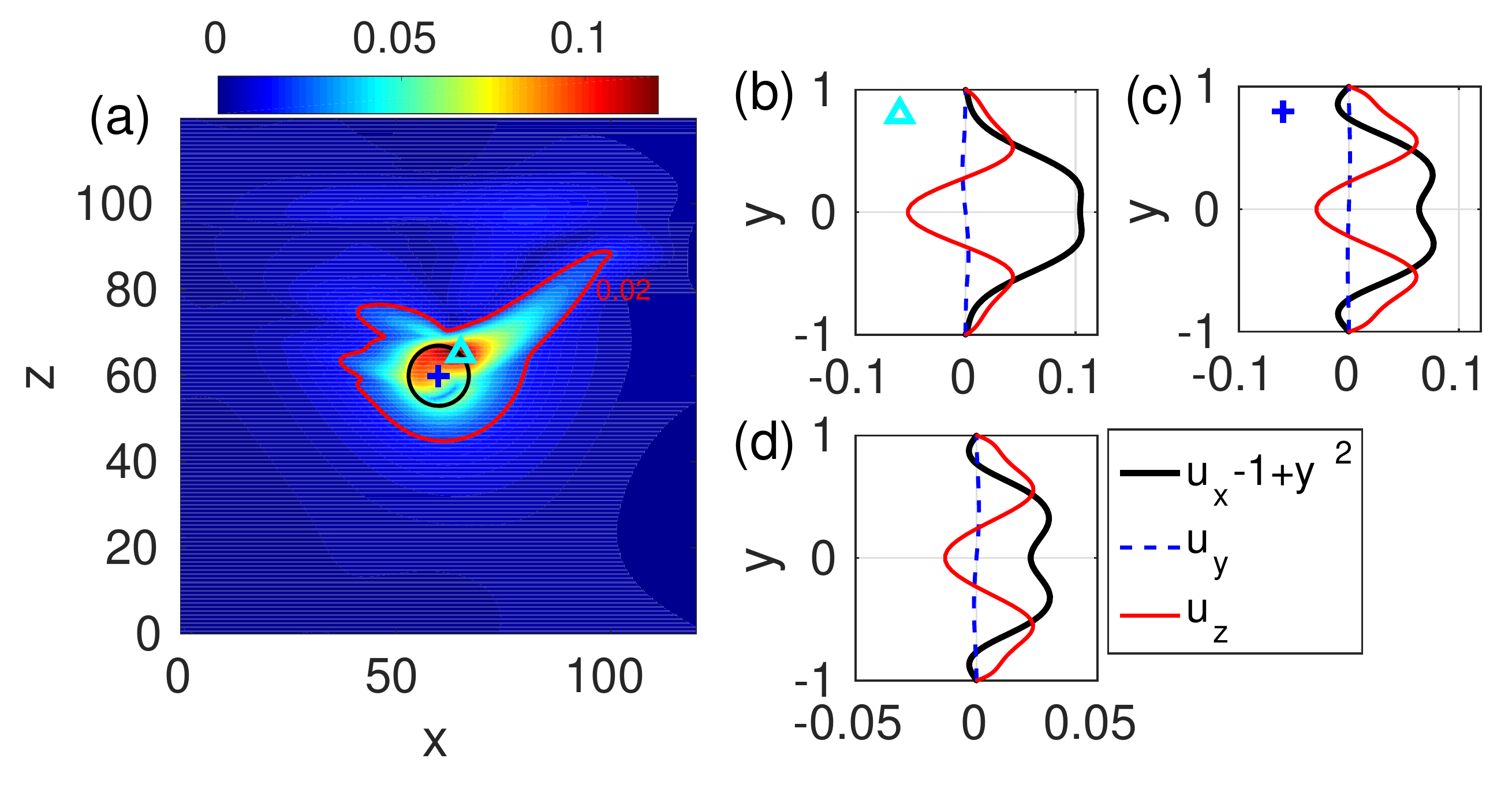}
\caption{\label{fig:effects_on_base_flow} The effect of the forcing on the base flow. $Re=750$ and $A_x=A_z=3.0$. (a) The maximum velocity deviation from the basic parabolic flow, $\max_y{\sqrt{(u_x-1+y^2)^2+u_y^2+u_z^2}}$, plotted in the $x-z$ plane. The black circle marks the forcing area at this time instant. \SB{The red line marks the level of 0.02.} (b-d) The velocity deviation profiles at the position marked by the cyan triangle (b), by the blue cross (c) and that averaged in the forcing area (inside the black circle) (d), i.e. \SB{$\frac{1}{N_p}\sum{u_{x,z}(y)}$, where $N_p$ is the number of grid points that satisfy $(x-x_c)^2+(z-z_c)^2<R^2$.}}
\end{figure}

\BS{To further evidence that the turbulent band is generated by the inflectional instability, we next show the effects of the forcing on the base flow for the $A_x=A_z=3.0$ case. To allow the base flow to develop sufficiently before the transition occurs, we performed a simulation without initial background noise, see Figure \ref{fig:KE_no_noise} and \ref{fig:effects_on_base_flow}. Comparing with the case with background noise (Figure \ref{fig:forcing_amplitudes}(a)), it takes approximately twice the time, roughly 600 time units, for the transition to occur. At $t=400$, at which the base flow seems to have stabilised under the forcing, we plotted the maximum velocity deviation with respect to the parabolic flow in the $x-z$ plane, see figure \ref{fig:effects_on_base_flow}(a), in which the forcing area is moving downward and to the right, and the black circle in the figure marks the forcing area at this time instant. Large velocity deviation is seen in the upper half and right shoulder of the forcing area. Although the velocity deviation is not limited in the forcing area, the region that the force significantly affects is not much larger than the forcing area \SB{(see the level of 0.02 marked in the figure)}, which means that the effect of the force is sufficiently localised and does not significantly influence the bulk and tail of the generated turbulent band. Panels (b) and (c) show the velocity profiles at the position of $(x,z)=(65,65)$ (marked by a cyan triangle in panel (a)), where the largest deviation is located, and at the center of the forcing area $(x,z)=(60, 60)$ (marked by a blue cross in panel (a)), respectively. Panel (d) shows the velocity profile averaged over $x$ and $z$ directions within the forcing area. Clearly, the force generates significant inflection in the local velocity profiles, which qualitatively resemble the velocity profile shown in Figure \ref{fig:target_profiles}. It should be noted that the forcing itself does not bring in turbulent fluctuations, see the stationary kinetic energy between $t=200$ and 600 in Figure \ref{fig:KE_no_noise}(a) and the nearly constant streamwise velocity at the center of the moving forcing area shown in panel (b). Besides, the magnitude of the velocity deviation induced by the force is much smaller than that of turbulent fluctuations inside turbulent bands. Figure \ref{fig:effects_on_base_flow} shows that the former is overall smaller than 0.1, whereas the maximum of the latter is no less than 0.3 in the cut plane at $y=-0.5$, see Figure \ref{fig:forcing_amplitudes}(d-e), and can be as large as around 0.5 in the whole domain, see Figure \ref{fig:compare_structure_Re600_750}(b,c). Roughly, it can be estimated that the kinetic energy density directly injected by the force is roughly one order of magnitude smaller than that associated with the generated turbulence. Therefore, the turbulent band is not directly fed by the force, instead, by the instability of the local flow induced by the force, see the wave-like streaks in Figure \ref{fig:forcing_amplitudes}(c-e), which locally resemble the unstable waves shown in the linear analysis (Figure \ref{fig:unstable_region_and_nonmodal_growth}(b)). 

\begin{figure}
\centering
\includegraphics[width=0.9\textwidth]{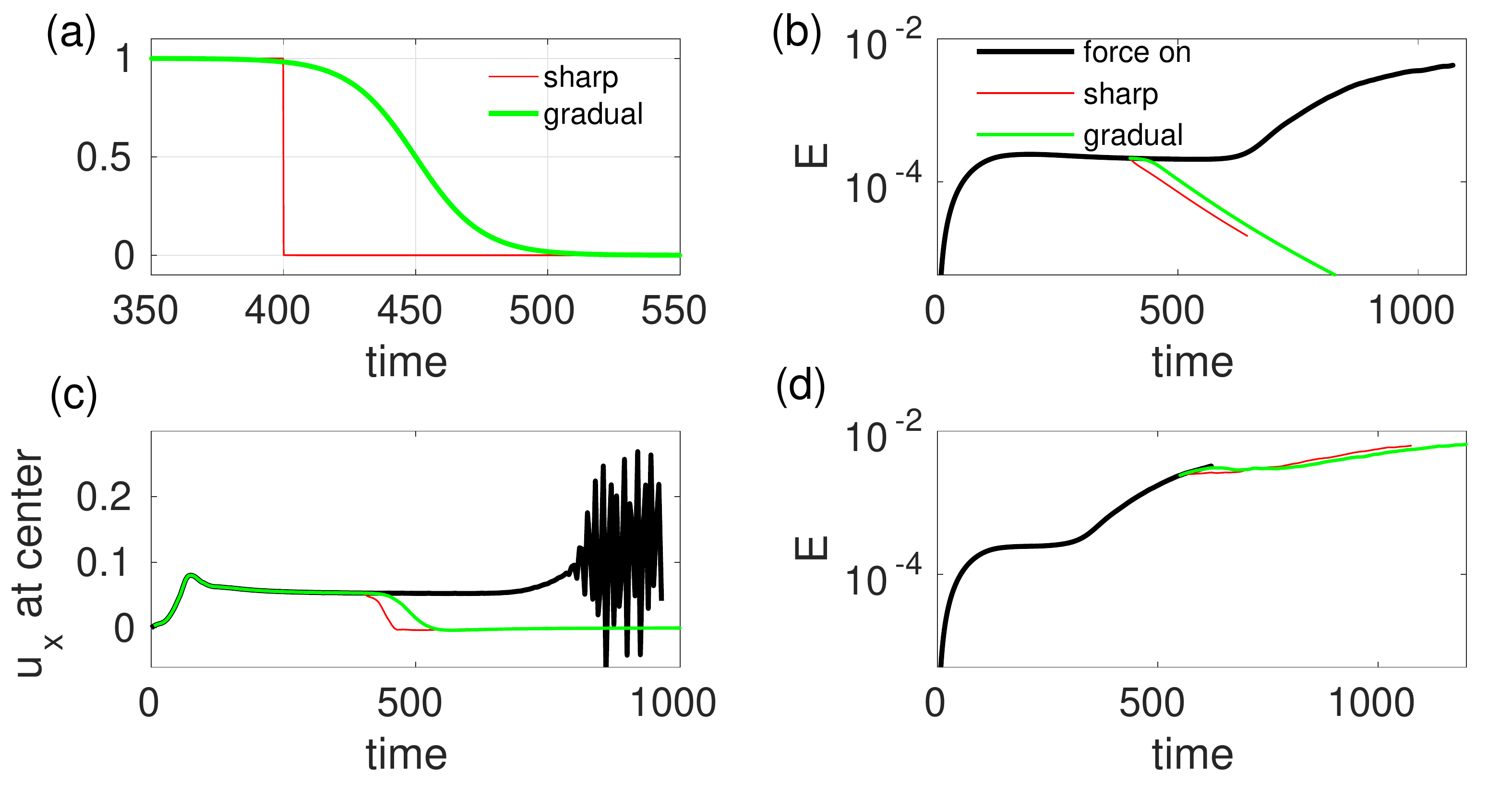}
\caption{\label{fig:different_deactivation} \SB{The effects of different deactivation of the force on the flow at $Re=750$. (a) The strength function of the sharp deactivation (thin red) and gradual deactivation (green) of the force tested in our study. (b) Deactivate the force according to (a) before the transition to turbulence occurs. 
(c) The streamwise velocity at the center of the forcing area in the simulations shown in (b). In panels (b,c), the bold black lines are taken from Figure \ref{fig:KE_no_noise}(a,b) and the sharp deactivation is at $t=400$. (d) Deactivate the force after the transition to turbulence occurs. The bold black line is the same as the one in Figure \ref{fig:forcing_amplitudes}(b). Note that the deactivation in (d) is the same as that in (b, c) except for a 150-time-unit delay (the sharp one is at $t=550$).}}
\end{figure}

\SB{In the results shown before, the force is turned off sharply. We also studied the effects of different deactivation of the force on the flow, see Figure \ref{fig:different_deactivation}. We compared a sharp deactivation and a gradual deactivation with the strength given by a tanh function, $0.5-0.5\text{tanh}((t-t_0)/25)$, where $t$ is time and $t_0$ is the nominal deactivation time instant, i.e. the instant when the strength halves. With this strength function, the force decreases smoothly to zero in approximately 100 time units. The sharp deactivation is at $t=400$ and the smooth one is at $t_0=450$ in panel (a). These two deactivation are applied to the simulation we have shown in Figure \ref{fig:KE_no_noise} to show the effect on the base flow. Panel (b) shows the kinetic energy of the flow, which immediately decreases after the force is removed or reduced without any sign of turbulent fluctuations. Panel (c) shows the streamwise velocity at the center of the forcing area in the $x-z$ cut plane of $y=-0.5$. Similarly, there is no sign of turbulent fluctuations induced by the deactivation. Therefore, the results indicate that the deactivation, no matter being sharp or gradual, does not cause any instabilities or turbulent velocity fluctuations, and therefore does not contribute to the generation of bands. Besides, we also tested the two deactivation after a short band has been generated. The simulation shown in Figure \ref{fig:forcing_amplitudes}(b) was considered. The sharp deactivation is at $t=550$, at which the short band generated by the force is visualised in Figure \ref{fig:forcing_amplitudes}(e). The smooth deactivation is at $t_0=600$ and the force starts to decrease approximately at $t=550$. Panel (d) shows the kinetic energy of the flow and that the band continue to grow in both deactivation cases. Therefore, these tests suggest that the deactivation of the force does not introduce instabilities and velocity fluctuations, and that the specific deactivation process does not significantly affect the band generation, as long as the band has sufficiently developed under the forcing. However, it can be noticed that the gradual deactivation slightly delays the growth of the band in terms of kinetic energy compared to the sharp deactivation case, see panel (d). A possible reason is that, the force does not contribute to the instability anymore as the strength gradually reduces, whereas it might slightly affect the formation of a natural local instability, or a natural head, that is needed for the turbulent band to grow. Therefore, we recommend a sharp deactivation for simplicity.}

To sum up, it is shown in this section that the forcing method we proposed generates local velocity profiles with similar properties as the target velocity profile (\ref{equ:target_profile}) (\ref{equ:target_profile_2}), which indeed can cause instability and trigger turbulent bands. } 

\subsection{Transient-band regime}

\begin{figure}
\centering
\includegraphics[width=0.95\textwidth]{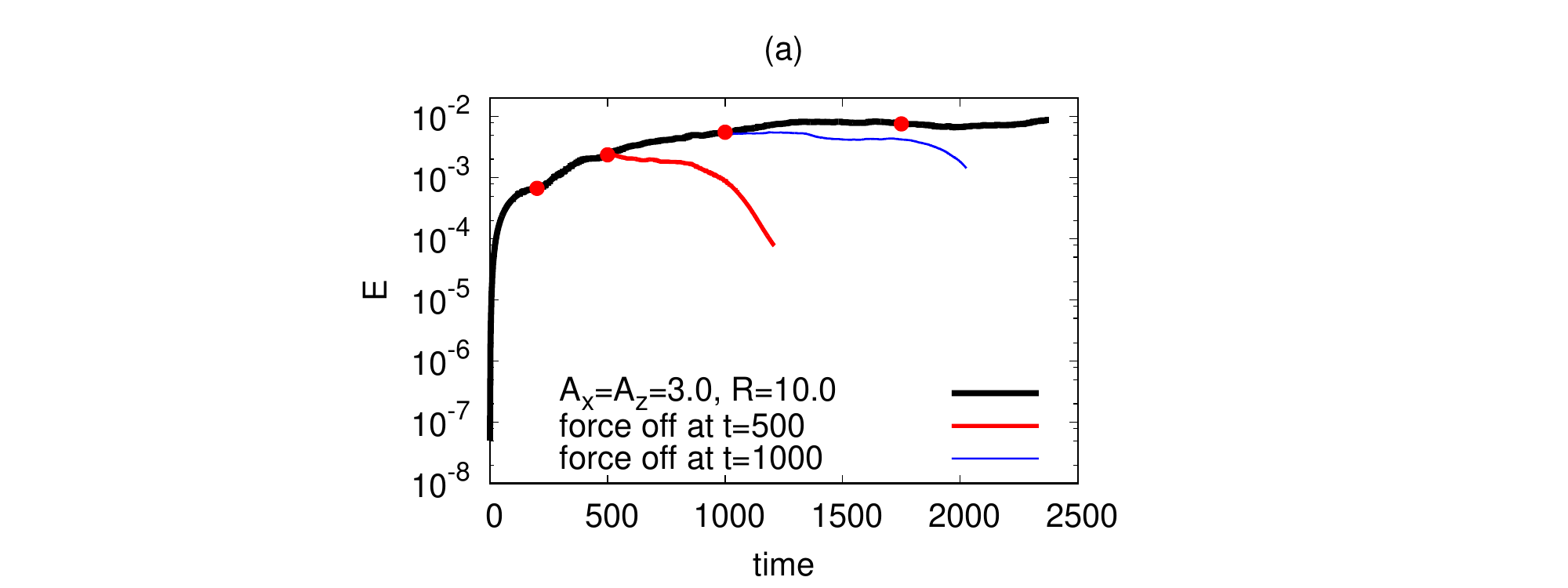}
\includegraphics[width=0.99\textwidth]{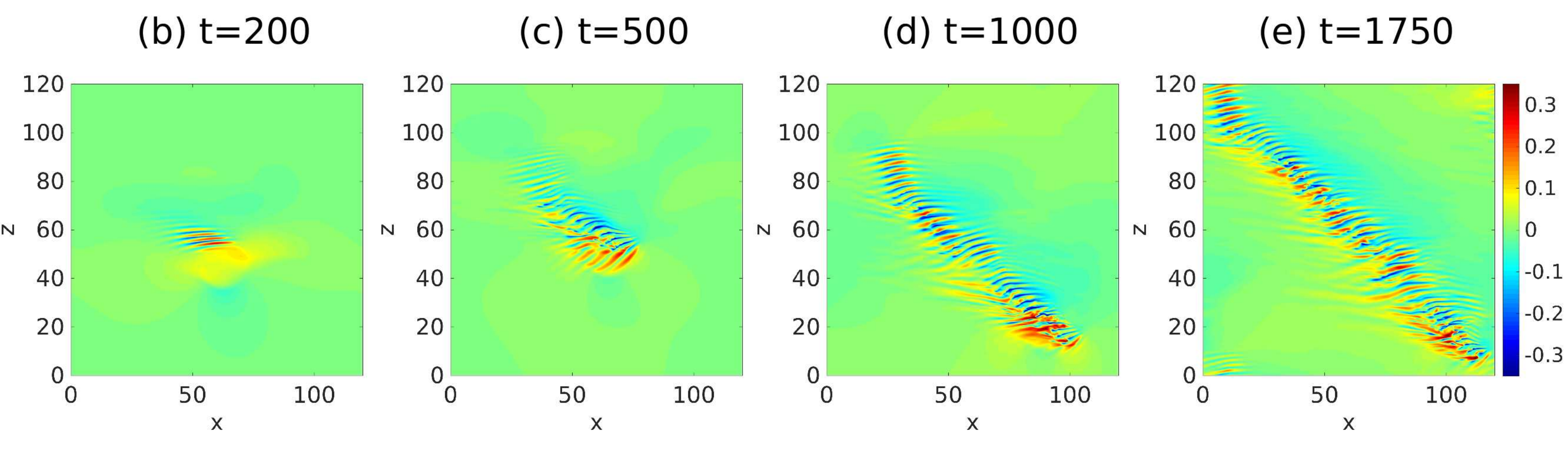}
\caption{\label{fig:forcing_Re600} (a) The kinetic energy of the velocity fields (with the parabola excluded) for $R=10.0$ and $A_x=A_z=3.0$ at $Re=600$. (b-e) Contours of the streamwise velocity in the $x$-$z$ cut plane at $y=-0.5$. Snapshots at time instants $t=200$, 500, 1000 and 1750 are shown (marked by red circles in panel (a)). Flow is from left to right.}
\end{figure}

In the previous section we showed that our method can generate turbulent bands at $Re=750$, which is in the sustained turbulent band regime. The band can sustain itself and grow even if the force is switched off. Here, we show results at Re=600 and 500, at which it has been shown that turbulent bands are unsustained \citep{Tao2018,Kanazawa2018,Shimizu2019}. 

We first performed simulations at $Re=600$ in the small box ($L_x=L_z=120$), and set $R=10.0$ and $A_x=A_z=3.0$. Figure \ref{fig:forcing_Re600} shows the kinetic energy of the flow (panel (a)) and the development of the turbulent band under the forcing (panel (b-e)). Clearly, the force successfully generates a turbulent band and sustains the band until it grows to the size of the computational domain (see panel (e)). We did not continue the simulation when the two ends of the turbulent band meet and the head starts moving into the band. However, starting from the flow field at $t=500$ (see panel (c)) and 1000 (see panel (d)), with the force deactivated, the turbulent band cannot sustain itself and eventually decays, indicated by the monotonically decreasing kinetic energy as shown by the red and blue thinner lines in panel (a). This indicates that turbulent bands are indeed unsustained at $Re=600$, in agreement with \citet{Tao2018,Kanazawa2018,Shimizu2019}. Nonetheless, the result seems to suggest that the band could be sustained if a turbulence-generating head, with a sufficiently strong spanwise inflection, could be maintained.

\begin{figure}
\centering
\includegraphics[width=0.95\textwidth]{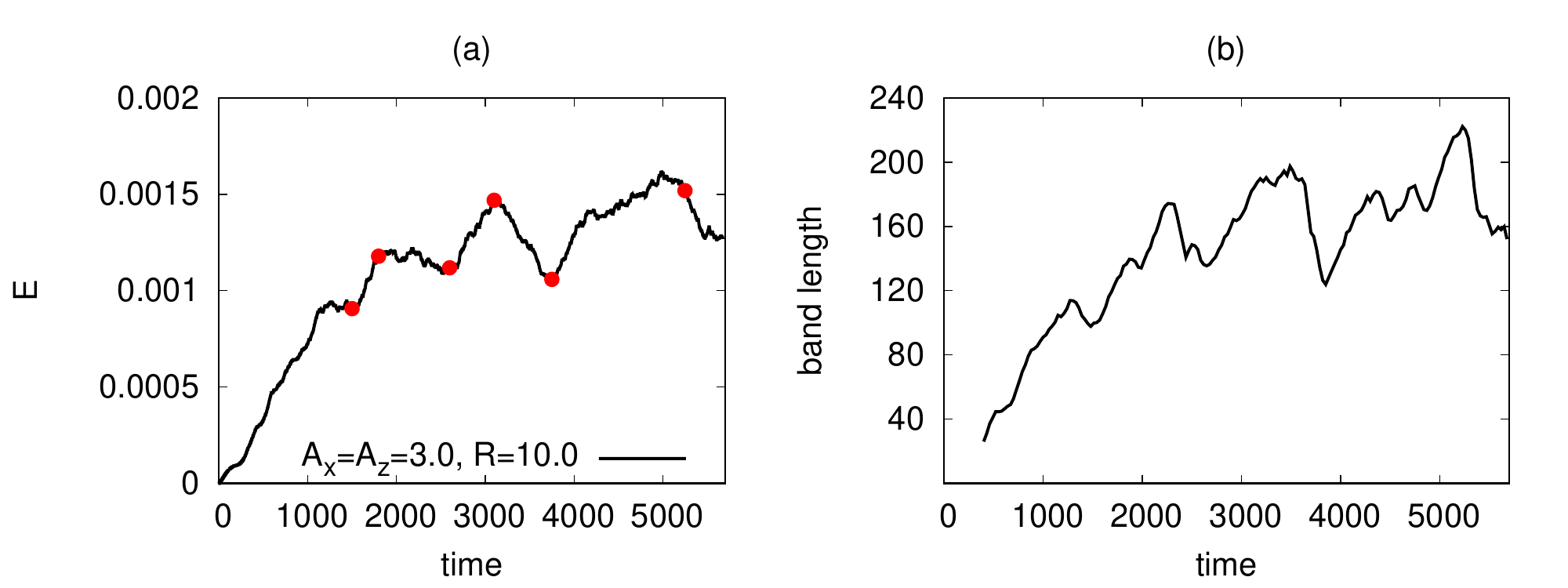}
\includegraphics[width=0.99\textwidth]{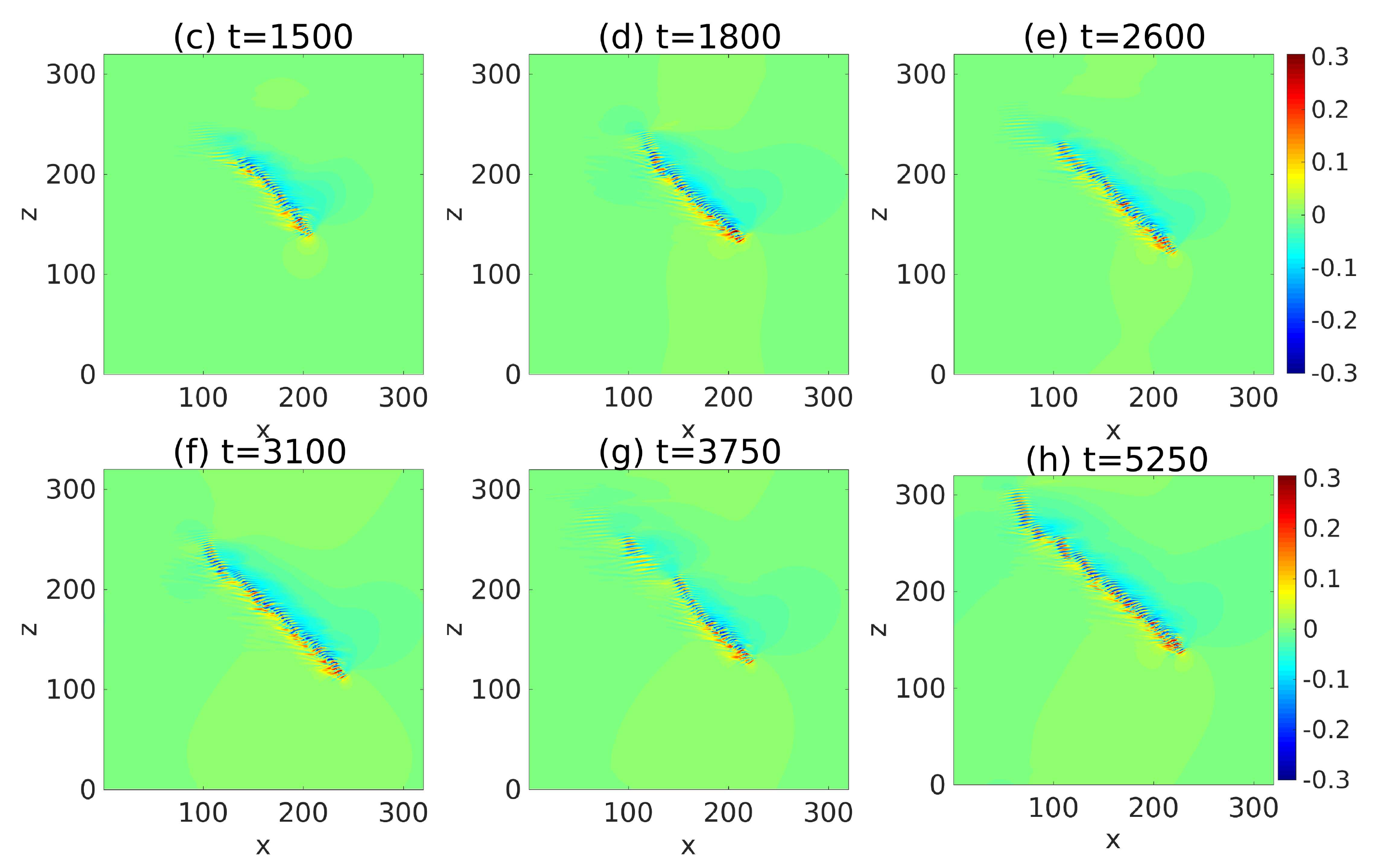}
\caption{\label{fig:forcing_Re600_large_domain} (a) The kinetic energy of the velocity field (with the parabola excluded) for $R=10.0$ and $A_x=A_z=3.0$ at $Re=600$. (b) The length of the band calculated using the method of \citet{Tao2018}. (c-h) Contours of the streamwise velocity in the $x$-$z$ cut plane at $y=-0.5$. Time instants $t=1500$, 1800, 2600, 3100, 3750 and 5250 are shown (marked by red circles in panel (a)). Flow is from left to right.}
\end{figure}

In the small domain, the band grows to the length of the diagonal of the domain under the forcing, which is approximately 170. In order to see if a turbulent band can persistently grow in length at such a low Reynolds number, if the restriction of the domain size is relieved, we also performed simulations in the large domain with $L_x=L_z=320$. Figure \ref{fig:forcing_Re600_large_domain} shows the development of the flow under the forcing. The simulation was performed up to 5700 time units and a turbulent band can grow to a length of around $200h$, see the length of the band calculated using the method of \citet{Tao2018} in panel (b) and the band at $t$=3100 in panel (f) and at 5250 in panel (h). However, 
the maximum length is not much larger than the final length of the band simulated in the small domain. 
The length shows large excursions \BS{(see panel (b))} due to that a long patch of the band at the tail decays continually (see Figure \ref{fig:forcing_Re600_large_domain}(f-h)), similar to the tail of sustained turbulent bands in large domains at $Re=660$ \citep{Kanazawa2018} and at $Re=750$ \citep{Xiao2020}. Our results seem to suggest that there is an upper bound for the length of the band at $Re=600$ (see the trend in Figure \ref{fig:forcing_Re600_large_domain}(b)), subject to the decay of the tail when the length of the band is sufficiently long, which is approximately $200h$ based on our data. The competition between the streaks generation at the head and the decay of the streaks at the tail should determine the length of the band. Unfortunately, it is too expensive to obtain statistics of the length of the band considering the very large computational domain and long observation time needed. 

\begin{figure}
\centering
\includegraphics[width=0.85\textwidth]{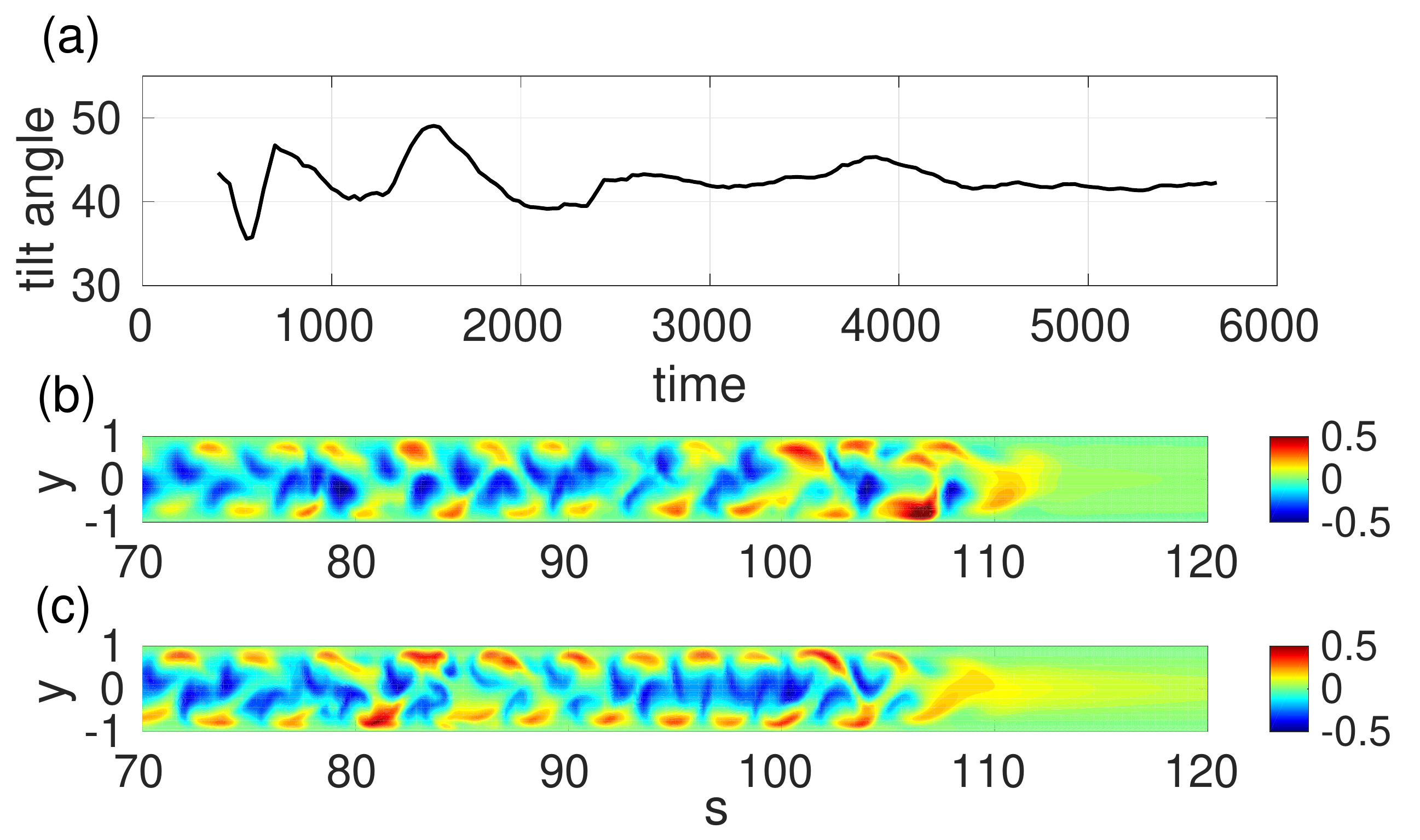}
\caption{\label{fig:compare_structure_Re600_750} \BS{The flow structure of the band at $Re=600$ as shown in Figure \ref{fig:forcing_Re600_large_domain}. (a) The tilt angle of the band calculated using the method proposed by \citet{Tao2018}. (b) The contours of streamwise velocity plotted in a cut plane along the band shown in Figure \ref{fig:forcing_Re600_large_domain}(d), in which $s$ denotes the coordinate along the band and $y$ is the wall-normal coordinate. Turbulent band is on the left hand side and laminar region is on the right hand side. The length in $y$ direction is stretched by a factor of 2 for a better display. (c) The same contour plot as in panel (b) for the sustained band at $Re=750$ shown in Figure \ref{fig:forcing_amplitudes}(f).}}
\end{figure}

\BS{Figure \ref{fig:compare_structure_Re600_750}(a) shows the tilt angle of the band shown in Figure \ref{fig:forcing_Re600_large_domain}, which fluctuates around $42^{\circ}$, excluding the large fluctuations at the early stage ($t<2000$). \citet{Kanazawa2018} reported that the average tilt angle of bands at $Re=660$ is approximately $41^{\circ}$, which is very close to what we found for $Re=600$. Panel (b) shows the flow structure between the two walls in a plane cutting through the band along its length, in which streamwise velocity is plotted as the colormap. Turbulent band is shown on the left hand side and the laminar region on the right hand side, and the head of the band is placed at around 110. We can see low speed streaks close to the channel center and high speed streaks close to the wall in the band. We chose the $x-z$ cut plane at $y=-0.5$ in our other figures because a plane too close to the channel center or too close to the wall would only cut through low speed or high speed streaks. For comparison, the same plot for a sustained band at $Re=750$ is plotted in panel (c). The structures in the bulk of the two bands (between 70 and 90) show high resemblance and that at the heads are also very similar even if the force is on for the $Re=600$ case while is absent for the $Re=750$ case. This wave-like alternating low and high-speed streaks pattern was also shown for a band at $Re=660$ by \citet{Kanazawa2018} (see Figure 5.2 in there) \SB{and was shown to be underlain by banded nonlinear exact solutions of the Navier-Stokes equations in channel geometry \citep{Paranjape2020}}. The similar tilt angles and flow structures between the band at $Re=600$ and the sustained ones at $Re=660$ and 750 suggest that the band we obtained in the transitional regime is not an artifact. A turbulent band could form and sustain if the inflectional instability mechanism at the head could be sustained. In other words, below $Re\lesssim 660$, turbulent bands are not sustained because the lacking of a sustained instability at the head.}

\begin{figure}
\centering
\includegraphics[width=0.95\textwidth]{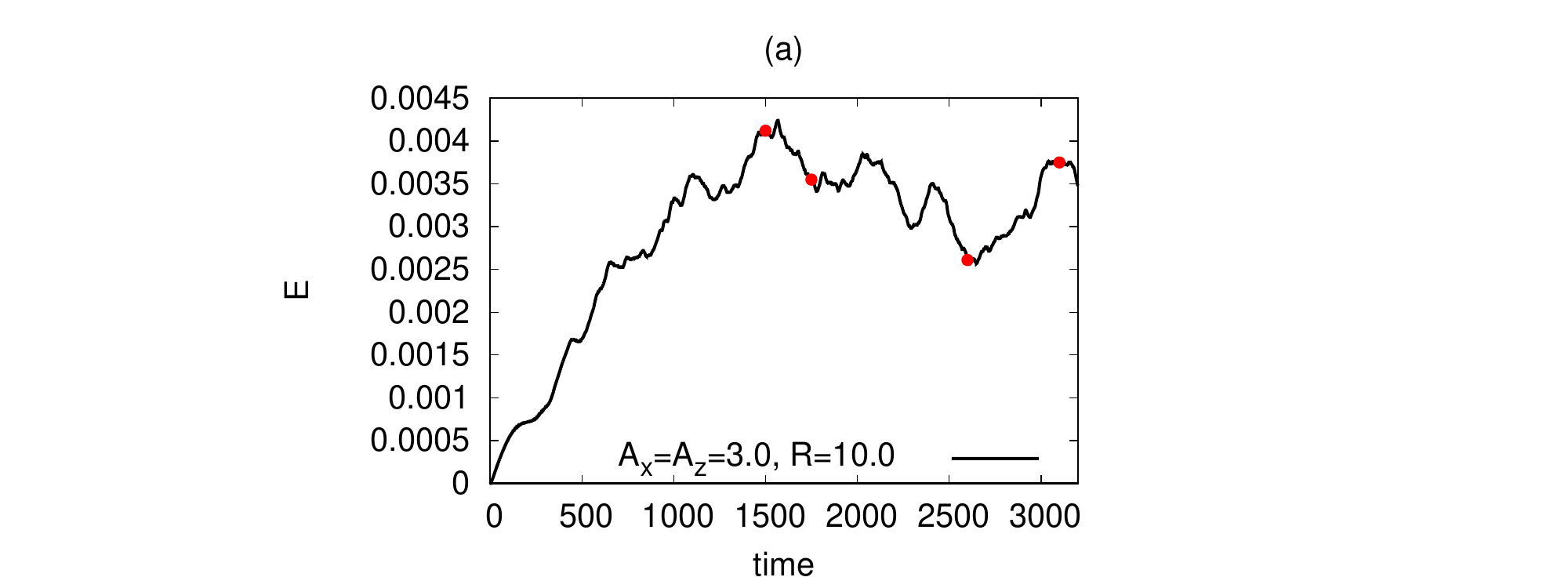}
\includegraphics[width=0.99\textwidth]{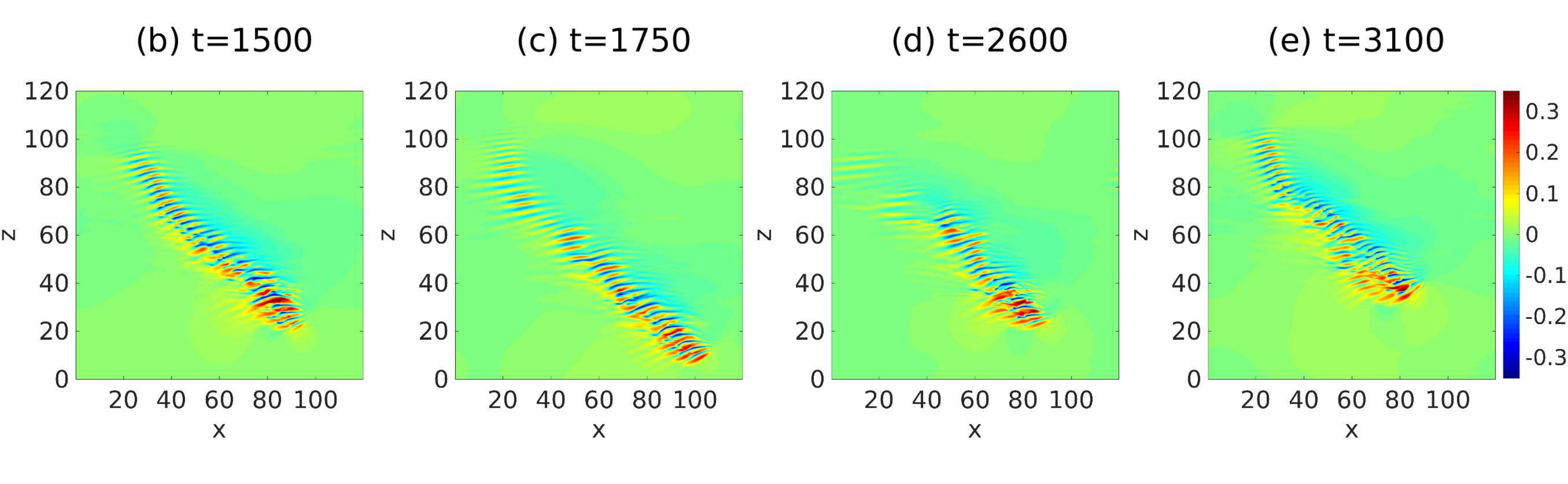}
\caption{\label{fig:forcing_Re500} (a) The kinetic energy of the velocity field (with the parabola excluded) for $R=10.0$ and $A_x=A_z=3.0$ at $Re=500$. 
(b-e) Contours of the streamwise velocity in the $x$-$z$ cut plane at $y=-0.5$. Time instants $t=1500$, 1750, 2600 and 3100 are shown.
}
\end{figure}

We went to further lower Reynolds numbers and performed simulations at $Re=500$. We chose the small computational domain and $A_x=A_z=3.0$ and $R=10$ for the force. Figure \ref{fig:forcing_Re500} shows that our method can still generate and sustain a band even at this low Reynolds number. Unlike the $Re=600$ case, the turbulent band only grows to a length of about $100h$ and the fast decay of the generated streaks \BS{strongly} limits the length of the band. Like the $Re=600$ case in the large domain, the band also undergoes a grow-shrink-grow cycle (see panel (b-e) in Figure \ref{fig:forcing_Re500}). As this small domain seems large enough for this Reynolds number, we did not consider the large domain.

\subsection{Effects of the size, steepness and speed of the forcing region}

\begin{figure}
\centering
\includegraphics[width=0.85\textwidth]{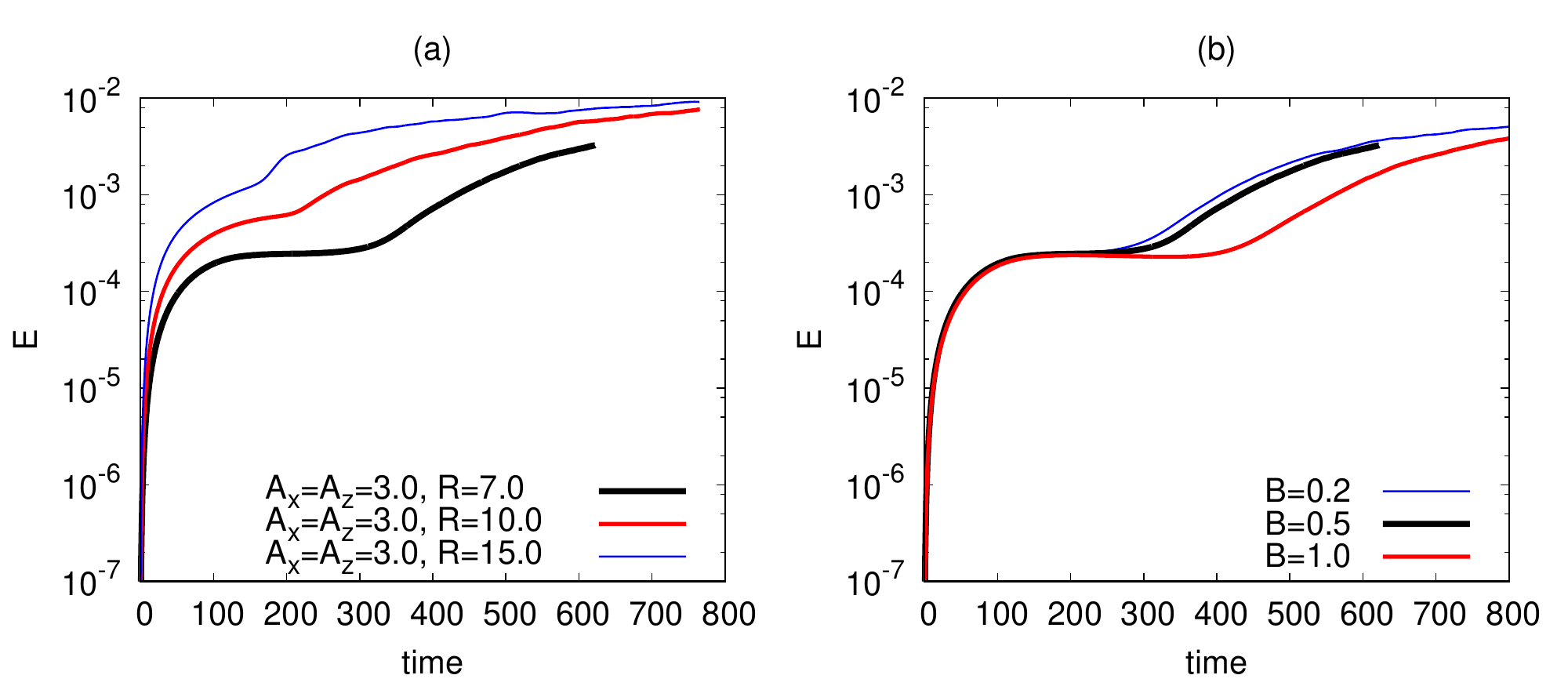}
\caption{\label{fig:forcing_area_size} (a) The effect of the size of the forcing area for the $Re=750$, $A_x=A_z=3.0$ and $B=0.5$ case. The kinetic energy of the velocity fields (with the parabola excluded) for $R=7.0$, 10.0 and 15.0 is plotted. \BS{(b) The effect of the steepness parameter $B$ for the $Re=750$, $A_x=A_z=3.0$ and $R=7$ case.}}
\end{figure}

The effect of the size of the forcing region was also briefly studied, see Figure \ref{fig:forcing_area_size}(a). We tested $R=7$, 10 and 15 and found that a larger forcing area can generate turbulent bands faster, given the same level of initial noisy perturbation (on the order of $\mathcal{O}(10^{-4})$). \BS{This is expected because, on the one hand, the base flow within the moving forcing region is exposed to the force for a longer time such that the inflectional velocity profile can develop more, and on the other hand, unstable waves can undergo a longer growth time, given the same moving speeds.} We did not consider further larger sizes as we intended to keep the size of the forcing area comparable with the width of the turbulent band. \BS{Fixing $A_x=A_z=3.0$ and $R=7$, we also tested the localisation factors $B=0.2$, 0.5 and 1.0. The results show that $B=1.0$ triggers the transition later than $B=0.2$ and 0.5, this is consistent with the effect of the size of the forcing region, because a smaller $B$ gives a steeper force curve at the boundary, which effectively gives a larger forcing area and can be clearly seen in Figure \ref{fig:localisation_factor}. However, as $B$ becomes small, the size of the forcing area is largely determined by $R$ and different $B$'s will give very similar results, as the $B=0.2$ and $0.5$ cases show in Figure \ref{fig:forcing_area_size}(b). However, we don't recommend very small $B$ because it may pose a strict restriction on the grid size.} 

\begin{figure}
\centering
\includegraphics[width=0.75\textwidth]{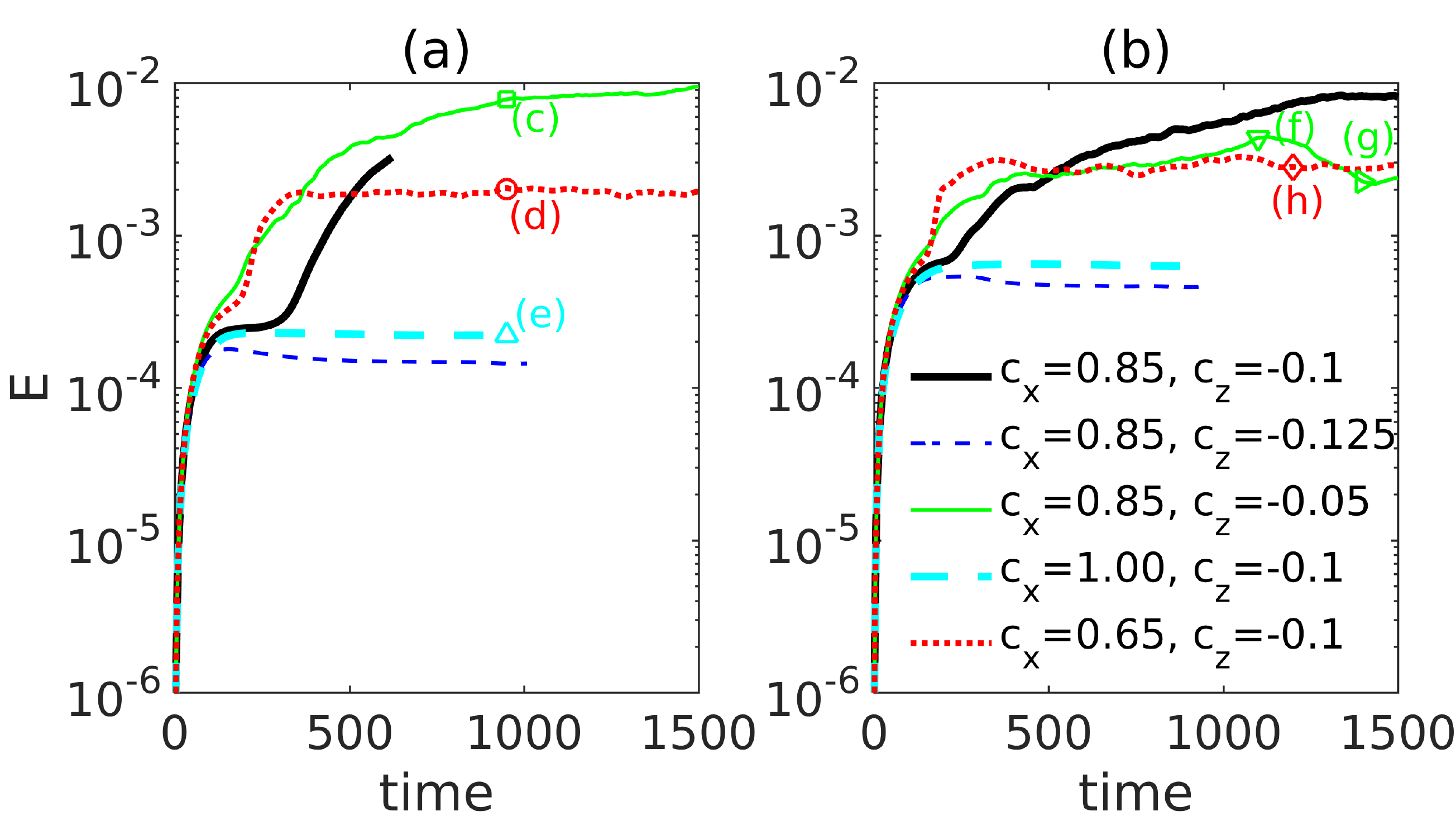}
\includegraphics[width=0.85\textwidth]{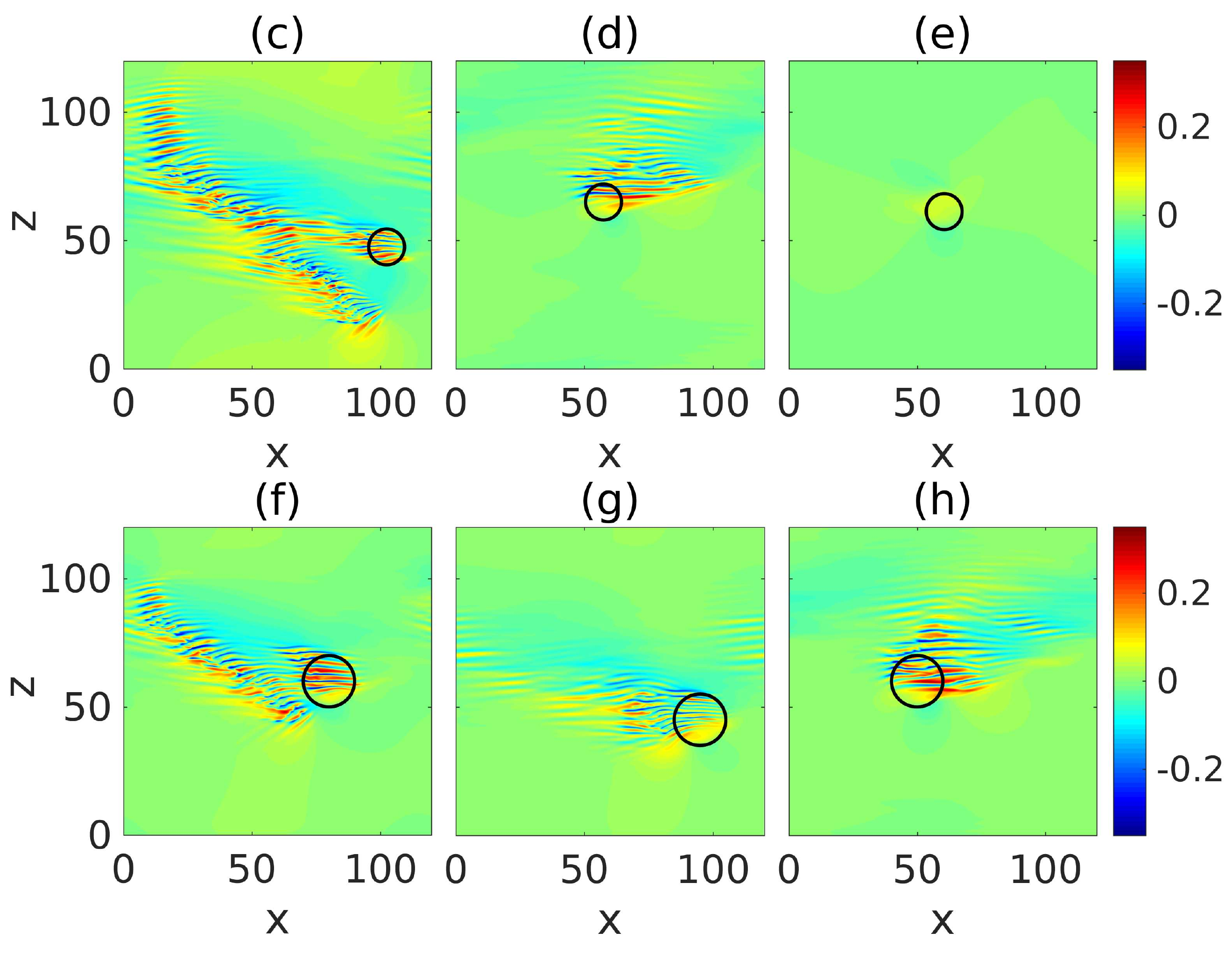}
\caption{\label{fig:speed_forcing_area} \BS{The effect of the speeds of the moving force for the $Re=750$ (a, c-e) and $Re=600$ (b, f-h) cases. The sizes of the forcing area are $R=7$ for the former and 10 for the latter, and $A_x=A_z=3.0$ for both cases. For each case, parameter groups of $(c_x, c_z)=(0.85, -0.05)$, $(0.85, -0.1)$, $(0.85, -0.125)$, $(0.65, -0.1)$ and $(1.0, -0.1)$ are considered. Panel (c-h) are contours of the streamwise velocity in the $x-z$ plane at $y=-0.5$ and are marked in panels (a, b) by the green square, red circle, cyan up-triangle, green down-triangle, green right-triangle and red diamond, respectively.}}
\end{figure}

\BS{In all the previous tests, the speed of the forcing area is set to be the natural speed of the head of turbulent bands at $Re=750$. In the following, we investigated the effect of the moving speed on the generation of turbulent bands. We changed one of the two speeds, i.e. streamwise and spanwise speeds, while keeping the other fixed, and considered the parameter settings of $(c_x, c_z)=(0.85, -0.05)$, $(0.85, -0.1)$, $(0.85, -0.125)$, $(0.65, -0.1)$ and $(1.0, -0.1)$. The results are shown in Figure \ref{fig:speed_forcing_area}. Panels (a) and (b) show the kinetic energy of the velocity deviations with respect to the parabolic flow at $Re=750$ and 600, respectively. The bold black line in the figure shows the baseline case with $(c_x, c_z)=(0.85, -0.1)$ as shown before in Figure \ref{fig:forcing_amplitudes} and \ref{fig:forcing_Re600}.

With the smaller spanwise speed (in absolute value) of $c_z=-0.05$, it is found that a band can still be triggered for both Reynolds numbers, see panels (c) and (f) in the figure. However, the forcing area moves more slowly than the band in the spanwise direction and gradually drifts away, see the black circles in panels (c) and (f). At $Re=750$, the band is sustained and continues to grow and the forcing area attempts to trigger another band when it is sufficiently far from the original band, see panel (c). However, a band is not sustained at $Re=600$ as we showed before. Indeed, the generated band gradually decays after the forcing area has drifted away for a certain time, see panels (f) and (g) which are separated by 300 time units as marked by the green triangles in panel (b). Presumably this generation-decay process will repeat if the simulation continues further. 

With the smaller streamwise speed of $c_x=0.65$, the force fails to trigger a turbulent band, although it still generates wave-like streaks, see panels (d) and (h) for $Re=750$ and 600, respectively. It can be seen that the wave-like streaks are mostly located outside the forcing area on the right hand side, suggesting that these structures are generated by the force but moves faster in the streamwise direction than the forcing area. The force generates a rather broad area of streaks which however do not develop into a band, see the saturated kinetic energy over a large time window shown by the red dotted line in panels (a, b). Besides, it is interesting to note that the instability occurs sooner with either smaller streamwise or spanwise speeds compared to the baseline case, see the earlier rapid increase in the kinetic energy shown by the red dotted and green thin lines in panels (a, b).

With either the larger streamwise speed of $c_x=1.0$ or spanwise speed of $c_z=-0.125$ (in absolute value), the force fails to trigger a band regardless of the Reynolds number, see the kinetic energy shown in panel (a) and (b). The visualisation of the case of $Re=750$ and $(c_x, c_z)=(1.0, -0.1)$ in panel (e) suggests that the force does not trigger an instability, at least not a sufficiently strong one, because no wave-like structures can be observed. There are two possible reasons. Firstly, the base flow does not have enough time to develop, therefore, cannot cause sufficiently strong instability, if the force moves too fast. Secondly, large mismatch between the speeds of the unstable waves and of the forcing region strongly limits the time window of the growth of the waves. Note that the wave-like streaky structures necessarily move slower than the natural speed of the head of the band, in both streamwise and spanwise directions, which can be inferred from the tilt direction of turbulent bands. This might also explain why sooner and significant instability can be triggered if the streamwise or spanwise speed of the force is smaller than the baseline case. However, we did not intend to investigate comprehensively the effect of all possible moving speeds. Instead, we propose the natural speeds we measured for $Re=750$ (the baseline speeds) as a protocol for cleanly generating a single band at low Reynolds numbers. In sustained band regime, a smaller spanwise speed may be used to speed up the formation of a turbulent band (see the green thin line in panel (a) and the baseline case). However, the force should be switched off at a proper time in order to avoid the forcing area drifting away and creating another band. In transient band regime, the baseline speeds are recommended if one wants to sustain a band for long times.
}

\section{Generation of pre-defined band patterns}
We show how to use the forcing strategy to generate a pre-defined band pattern precisely and then study various interactions between bands. The former results only showed a single tilt direction of the band. In fact, with our method, one can easily generate a band with the opposite tilt direction. The only thing one needs to do is to reverse the sign of the spanwise force and the sign of the spanwise speed of the moving forcing area, i.e., $f_z$ and $c_z$. With a precise control on the tilt direction and position of the band, we can study complex interactions of turbulent bands at low Reynolds numbers.

\begin{figure}
\centering
\includegraphics[width=0.95\textwidth]{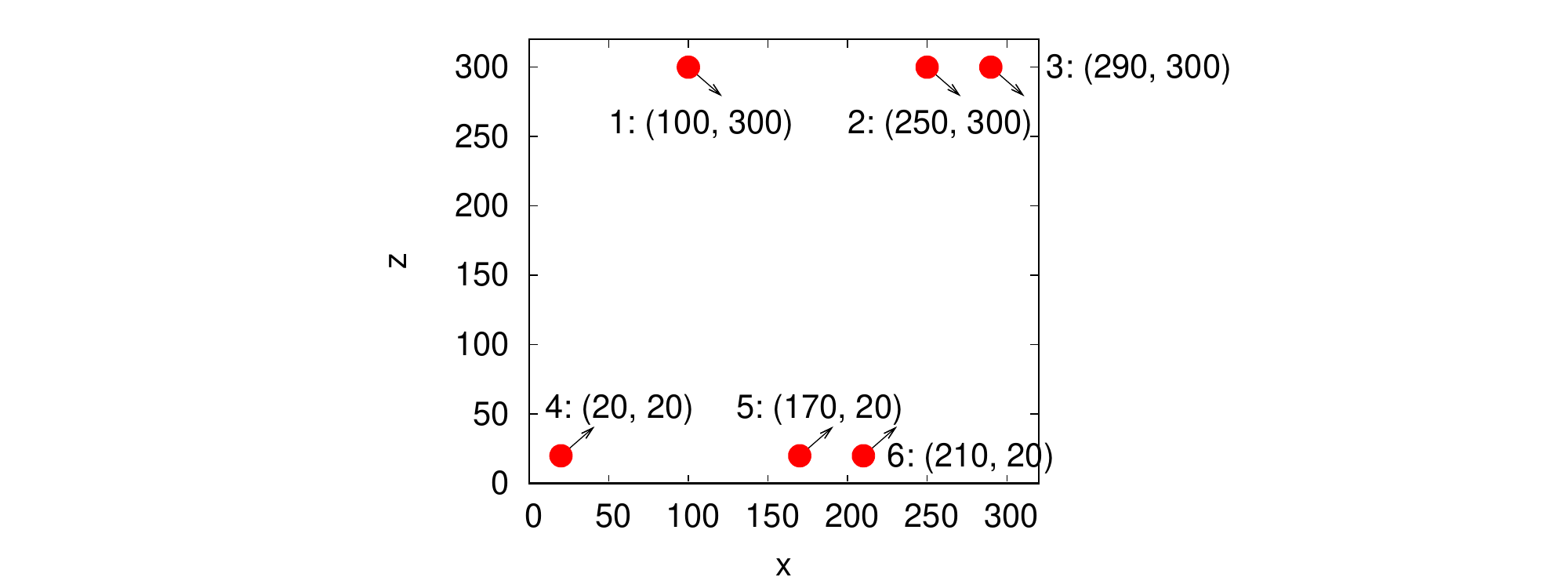}
\caption{\label{fig:six_circles_arrows} The initial distribution of forces in the large domain. Six forces are considered, labeled by 1 to 6. The initial position of the forces are also shown. The arrows roughly shows the moving direction of the forcing area. $Re=750$, $A_x=A_z=3$ and $R=7$ are considered. For forces 1-3, $c_z=-0.1$ and for forces 4-6, $c_z=0.1$, whereas $c_x=0.85$ for all forces.
}
\end{figure}

For this study, we considered the large domain. Figure \ref{fig:six_circles_arrows} shows the initial positions and moving directions of the forces. With six forces, we aimed to generate six turbulent bands. Bands generated by forces 1-3 (referred to as bands 1-3 hereafter) move in the same direction and have the same orientation with respect to the streamwise direction and bands 4-6 move towards bands 1-3 with the opposite orientation compared to bands 1-3. These two groups of forces are separated by a distance of 280 in the spanwise direction (for detailed positions of the forces, see Figure \ref{fig:six_circles_arrows}), which offers sufficient time for the bands to form under the forcing. With this forcing pattern, we wanted to investigate at least two types of interaction between turbulent bands at the same time. One is the longitudinal interaction when two parallel bands are located closely, such as the band pairs (2, 3) and (5, 6), and the other is the collision of two bands when two bands with opposite orientations intersect, which is expected to happen between band pairs (1, 4), (1,5), (5, 2) etc.

\begin{figure}
\centering
\includegraphics[width=0.9\textwidth]{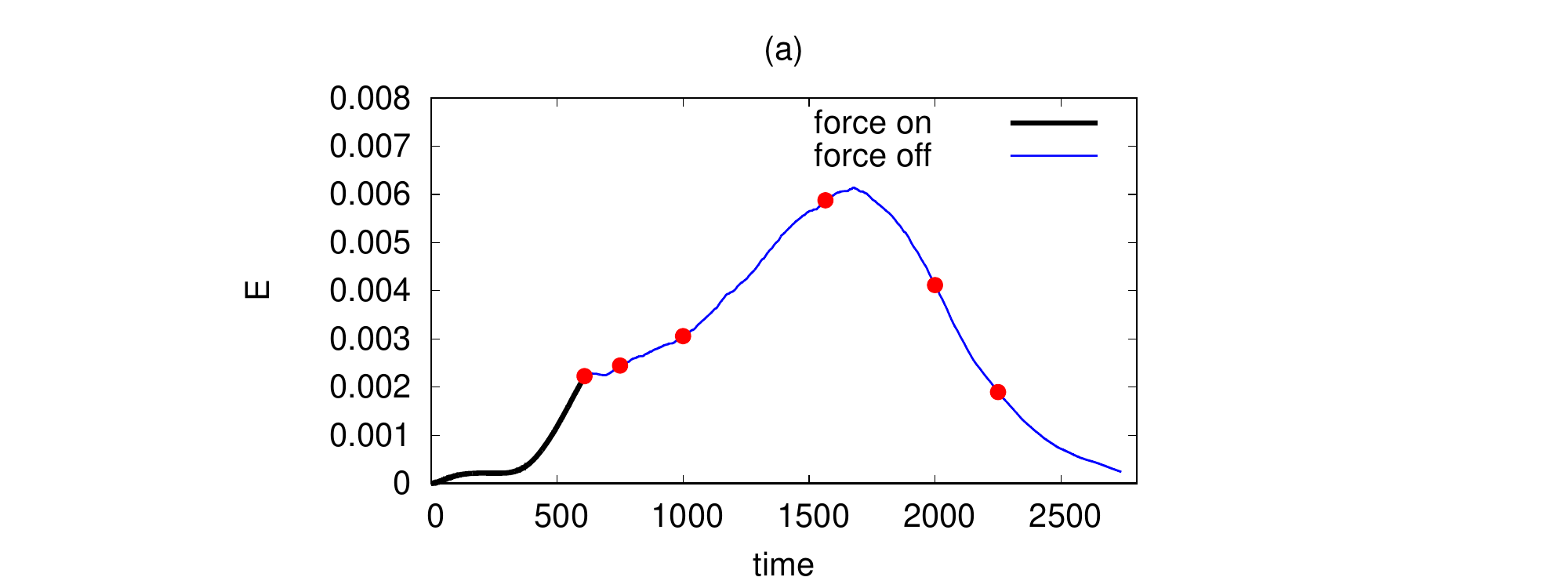}
\includegraphics[width=0.99\textwidth]{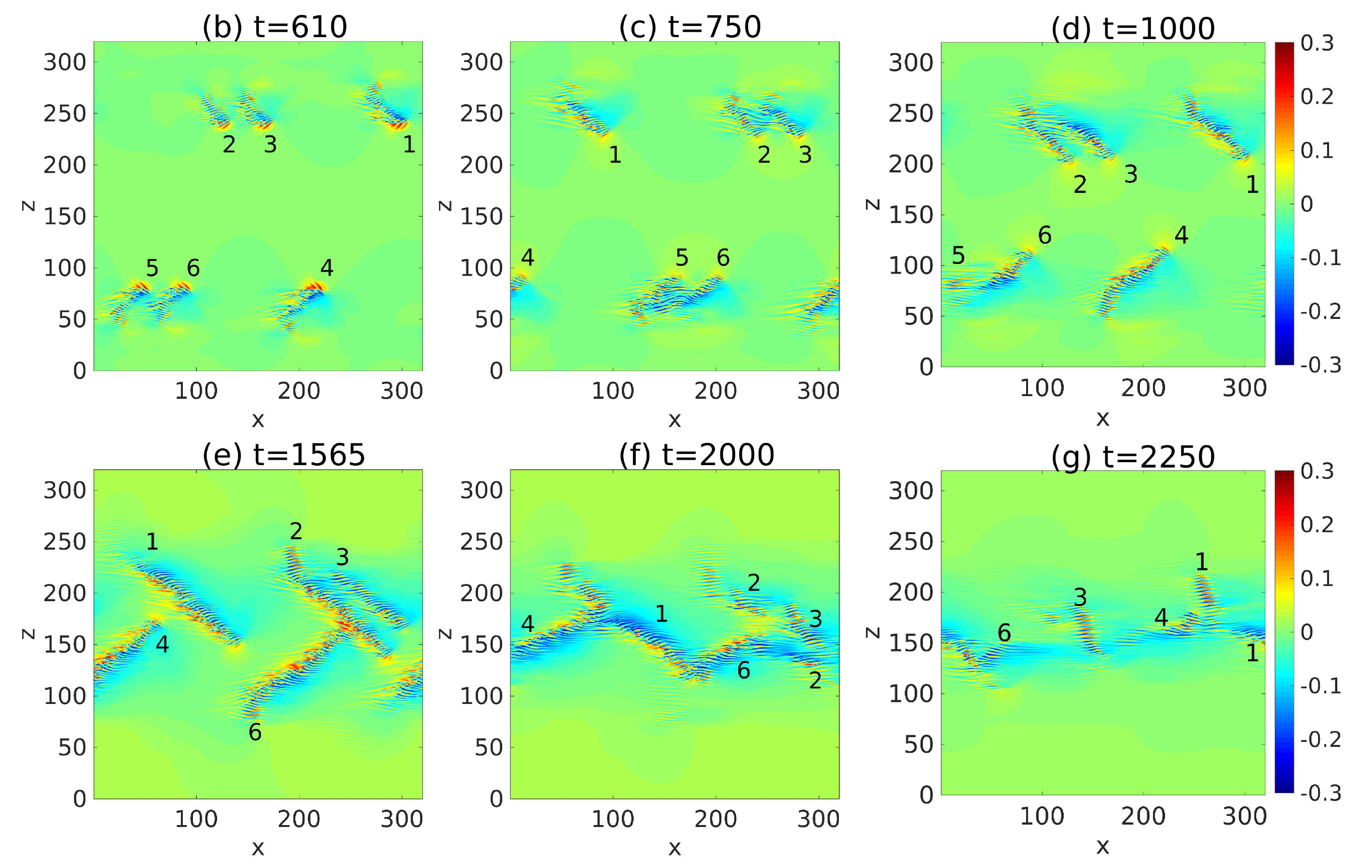}
\caption{\label{fig:Re750_multi_forces} (a) The kinetic energy of the velocity field (with the parabola excluded) at $Re=750$. Forces with $R=7.0$ and $A_x=A_z=3.0$ are imposed until $t=610$. (b-e) Contours of streamwise velocity in the $x$-$z$ cut plane at $y=-0.5$. Time instants $t=610$, 750, 1000, 1565, 2000 and 2250 are shown (marked by red circles in panel (a)). Bands generated by forces are labeled according to the label of forces as shown in Figure \ref{fig:six_circles_arrows}.
}
\end{figure}

Figure \ref{fig:Re750_multi_forces} shows the kinetic energy of velocity fluctuations and visualises the development of the flow. Panel (b) shows that six bands, though still short, are successfully triggered at $t=610$ with our forces. Subsequently, the forces are deactivated and the bands are sustained at this Reynolds number and continue to grow. The bands visualised in panel (c) at $t=750$ show that they have the characteristics of typical turbulent bands at low Reynolds numbers, i.e., an active streak generating head at the downstream end (see the red wave-like streaks at the downstream end) and a weak diffusive tail at the upstream end. This indicates that the bands are not artifacts due to the forces but actually self-sustained. It can be seen that bands 5 and 6 nearly overlap due to the small separation, so do bands 2 and 3 but with seemingly weaker interaction. As the flow develops, band 5 and 6 strongly interact which results in the decay of band 5, see panel (d) at $t=1000$. \SB{As \citet{Tao2018} proposed, a band relies on the large scale secondary flow surrounding the band, and a close neighbour may affect this secondary flow and eliminate the band. \citet{Xiao2020} speculated that the inflectional mean flow at the head of the band is sustained by this secondary flow. 
Alternatively, \citet{Kanazawa2018} proposed that the head of turbulent bands is locally self-sustained and does not depend on the bulk of the band. In either senario, the secondary surrounding flow of a close band may affect the head such that the instability is weakened or even eliminated, leading to the decay of the band. However, the exact self-sustaining mechanism of the head of turbulent bands is still poorly understood. Resolving this mechanism is the key to unravelling the parallel interaction between bands.} The rest of bands keep growing and at around $t=1565$ (panel (e)), bands 6 and 2 collide and the streak generating head of band 6 is also destroyed. Bands (1, 4), bands (2, 4) and bands (1, 6) are also close to a collision. It should be noted that band 3 has already started to decay because of the interaction with band 2, which is indicated by the disappearance of a streak generating head (to compare with other bands) at the downstream end, similar to what happened between band 5 and 6. Later, as collisions occur, the total kinetic energy of the flow field starts to quickly decrease, see panel (a). This indicates that once the streak-generating head is lost, turbulent bands cannot be sustained and will decay. The results support 
\BS{the conclusion of \citet{Kanazawa2018}, \citet{Shimizu2019} and \citet{Xiao2020} that turbulent bands at low Reynolds numbers are driven by a streak-generating end.} At $t=2000$, band 2 and 3 have nearly completely disappeared. Band 1, 4 and 6 are also decaying due to the collision. All bands have nearly decayed at $t=2250$ and the flow continues to relaminarise without any recovery of turbulence, see the kinetic energy shown in panel (a). \BS{Similar parallel elimination and annihilation upon collision between bands were also reported by \citet{Shimizu2019}}.

\section{Discussion and conclusion}
Inspired by the recent work of \citet{Xiao2020} which proposed that turbulence generation at the head of turbulent bands is driven by a spanwise inflectional local mean flow via a linear instability, we developed a perturbation method which can trigger turbulent bands directly at very low Reynolds numbers. This method features imposing a body force that induces a local flow with a sufficiently strong spanwise inflection. The force is designed using a fit of the velocity profile locally measured at the head of a turbulent band at $Re=750$ by \citet{Xiao2020}. We revisited the stability properties of the profile and showed fast non-modal growth associated with the profile. The non-modal growth analysis shown here complements the modal analysis of \citet{Xiao2020} and reinforces our belief that a turbulent band is driven by a local mean flow with a strong spanwise inflection via linear mechanisms. 

The moving force method \BS{can effectively trigger an instability which subsequently generates turbulent bands at low Reynolds numbers. In the sustained band regime, the force can be switched off once the band has sufficiently developed. In the unsustained band regime below $Re\simeq 660$,} we can also trigger and maintain turbulent bands without switching off the force. Our simulation at $Re=600$ shows that turbulent bands can grow to a length up to around 200$h$ under the forcing at the head. When the band reaches this length, the decay of the tail of the band seems to balance the streak generation at the head, and this dynamic balance keeps the band from growing further longer. \BS{That turbulent bands do not grow persistently in length at low Reynolds numbers has also been reported in the sustained band regime by \citet{Kanazawa2018}, who showed an average length of roughly 300$h$ for a sustained band at $Re=660$.} In contrast, if the forcing is deactivated, as shown in Figure \ref{fig:forcing_Re600}, a turbulent band cannot sustain itself and decays. These results suggest that, if the local mean flow at the head could maintain a sufficiently strong spanwise inflection, the band could be sustained even at $Re=600$, \BS{and turbulent bands are not sustained at $Re\lesssim 660$ because of the lacking of a naturally sustained inflectional instability at the head.} The results at $Re=500$ are similar, but turbulent bands cannot grow very long due to the very fast decay of streaks at the tail. The successful generation of bands at very low Reynolds numbers using the method in turn supports the turbulence generation mechanism of turbulent bands that we proposed here and in \citet{Xiao2020}. 

An important feature of this method is that it enables to generate turbulent bands with pre-defined positions and orientations in numerical simulations. \BS{According to our knowledge, there have been no such methods reported} in the literature. As shown in Figure \ref{fig:Re750_multi_forces}, we designed such a case at $Re=750$. 
We indeed observed very interesting annihilation of bands when collisions between bands with opposite orientations occur and elimination when two parallel bands are located closely to each other, see Figure \ref{fig:Re750_multi_forces}. Very interestingly, the six bands eventually all decay due to the special initial positions and orientations. Our study clearly showed that once the head of a band is destroyed, the band cannot sustain and decays even in the sustained band regime. 

Similar interactions between bands were reported by \citet{Shimizu2019}. The authors showed that the interaction determines the final flow pattern at the equilibrium state. At low Reynolds numbers, turbulence cannot form two-sided band pattern but can only form parallel turbulent bands. Our results agree with their finding. Further, they speculated that the directed percolation-like behaviour only starts to emerge when the two-sided band state is reached. However, further quantitative studies on various types of interaction between bands are need to elucidate the transition at low Reynolds numbers, prior to the onset of directed percolation. We believe that a precise control on relative positions and orientations of turbulent bands is important for relevant studies.

So far we have only shown the interactions between turbulent bands in a periodic channel without side walls. However, in laboratory experiments, periodic boundary conditions cannot be realised and the spanwise motion of turbulent bands means that they have to meet channel side walls, if sufficiently long observation time is desired. When a turbulent band gets close to the side wall, certainly the interaction between the band and the side wall will occur. Considering the important role that the streak-generating head of the band plays in the self-sustaiment of the band, the interaction between the head and the wall could significantly affect the self-sustainment of the band, \BS{which would certainly be interesting for experimental studies} and will be our further study.

\section{Acknowledgements}
We acknowledge financial support from the National Natural Science Foundation of China under grant number 91852105, 91752113 and from Tianjin University under grant number 2018XRX-0027. We acknowledge the computing resources from National Supercomputer Center in Guangzhou.

\section*{Declaration of interests}
The authors report no conflict of interest.


\end{document}